\documentclass[conference]{IEEEtran}

\usepackage{subcaption}
\usepackage{mathptmx}
\usepackage{amsmath}
\usepackage{multicol}
\usepackage{wrapfig}
\usepackage{array}
\usepackage{amssymb}
\usepackage{tikz}
\usetikzlibrary{arrows,shapes,chains,matrix,positioning,scopes,patterns}
\usepackage[linesnumbered,ruled,vlined]{algorithm2e}
\usepackage{graphicx}
\usetikzlibrary{decorations.pathreplacing}
\graphicspath{{img/}}
\title{Adaptive Content-based Routing using Subscription Subgrouping in Structured Overlays}
\author{Muhammad Shafique \\ School of Computer Science, University of Waterloo\\alummi@acm.org}
\begin{document}        
\maketitle
\setlength\abovedisplayskip{0pt}
\setlength\belowdisplayskip{0pt}

\tikzstyle{line} = [draw, -latex']
\begin{abstract}
Cyclic or general overlays may provide multiple paths between publishers and subscribers. However, an advertisement tree and a matching subscription activates only one path for notifications routing in \textit{publish/subscribe} systems. This poses serious challenges in handling network conditions like congestion, and link or broker failures. Further, content-based dynamic routing of notifications requires instantaneous updates in routing paths, which is not a scalable option. This paper introduces a clustering approach with a bit-vector technique for inter-cluster dynamic routing of notifications in a structured cyclic topology that provides multiple paths between publishers and interested subscribers. The advertisement forwarding process exploits the structured nature of the overlay topology to generate advertisement trees of length 1 without generating duplicate messages in the advertisement forwarding process. Issued subscriptions are divided into multiple disjoint subgropus, where each subscription is broadcast to a cluster, which is a limited part of the structured cyclic overlay network. We implemented novel static and \textit{intra-cluster} dynamic routing algorithms in the proposed overlay topology for our advertisement-based publish/subscribe system, called \textit{OctopiA}. We also performed a pragmatic comparison of our two algorithms with the state-of-the-art. Experiments on a cluster testbed show that our approach generates fewer inter-broker messages, and is scalable. 
\end{abstract}
\section{Introduction}
Distributed \textit{publish/subscribe (PS)} systems are used an notification service for many-to-many communication among distributed applications \cite{MANY_FACES}. Due to offering decoupled nature of interaction, PS systems are widely used in workflow management \cite{work_flow}, business process execution and monitoring \cite{muthy_thesis}, massive multi-player on-line games \cite{MMOG_Canas}, software defined networking \cite{TariqPLEROMA}, etc. PS systems are also used in a number of commercial applications. For example, Yahoo uses PS in PNUTS \cite{PNUTS}; Google has introduced Google Cloud PS \cite{G_CLOUD_PS}; Microsoft uses Bing Pulse  \cite{MS_PULSE}; LinkedIn uses Kafka \cite{KAFKA}; and Wormhole handles messages generated by Facebook users \cite{WormHole}. In enterprise environments, publish/subscribe also used in different contexts and standards including algorithmic trading \cite{algo_trading}, WS-notification and eventing, RSS feeds \cite{Cobra}, Enterprise Service Bus (ESB) \cite{ESB}, and Internet of Things (IoT) \cite{IoT}, etc.

In mostly PS systems, a set of \textit{brokers} connect each other to form \textit{overlay network} to provide notification service. Data senders (a.k.a. \textit{publishers}) connect to a nearby broker to publish the data in form of \textit{notifications}, while receivers (a.k.a. \textit{subscribers}) register their interest in the form of \textit{subscriptions} to receive notifications of interest \cite{carz_thesis,PADRESBookChapte}. Publishers and subscribers are decoupled in time, space, and flow, and remain anonymous to each other \cite{MANY_FACES}. In \textit{advertisement-based PS (APS)} system, an \textit{advertisement} is issued by a publisher to declare the set of notifications it is going to send. Each advertisement is broadcast to an overlay network and saved by each broker in its \text{Subscrption Routing Table (SRT)} to form an advertisement tree. The root of the tree is the broker which hosts the advertisement issuing publisher. Advertisements and subscriptions are normally composed of one or more filters, which are constraints on types and contents of notifications. An advertisement matches with a subscription if both have a non-empty overlap, which indicates that the subscriber that issued the subscription is interested in receiving notifications from the publisher, which issued the matching advertisement. Notification routing paths are created from the host broker of a subscriber to host brokers of publishers that issued matching advertisements. A subscription is forwarded, in reverse, to only those brokers, which are nodes of the matching advertisements trees. The subscription is saved in \textit{Publication Routing Table (PRT)} of each broker in a routing path. At each broker in the routing path, a notification is routed stepwise to next broker after matching contents of the notification with interested subscriptions. This notification routing technique is called the \textit{content-based} or \textit{filter-based} routing using \textit{Reverse Path Forwarding (RPF)} \cite{RPF}\cite{SIENA_WIDE_AREA}. APS systems are used as notification service when the expected number of subscribers are higher than the number of publishers. APS systems reduce the number of entries in PRTs of brokers and also improve the matching process for sending notifications to interested brokers \cite{PADRESBookChapte}. As publishers announce their advertisement before sending notifications, detection of \textit{patterns} in an APS systems is much efficient than the subscription-based variant of a PS system \cite{composite_pat}. Further, APS systems are suitable for publisher-offered and subscriber-requested provisioning of quality of service in content-based routing \cite{IndiQoS}.

Most APS systems use acyclic topology and offer only one routing path for sending notifications (ref-2010-survey-paper). Routing is made simple by acyclic overlays; however, a single routing path may have an unpredictable number of brokers without local interested subscribers. This condition causes inefficient use of network resources because a number of brokers in notifications routing paths act as \textit{forwarder-only} brokers and generate \textit{pure forwards} in large overlay networks \cite{reza-thesis}. \textit{Dynamic Clustering} is used as another approach to increase system throughput \cite{Com_clustering}. Subscribers with similar interests are moved to brokers in close proximity to generate less number of replicas of notifications. However, in Internet-scale APS systems, where subscribers have widely varying interests and connect from different geographical regions, it is difficult to attain an optimal performance through clustering because finding subscribers with similar interests is a costly process. PS systems that use acyclic overlays have limited flexibility for dealing with varying network conditions. For example, they are not robust in handling situations like load imbalance, link congestion, and broker or link failure. In load balancing, several \textit{coordination messages} have to be exchanged between brokers to select a broker for load shifting \cite{dynamic_LoadBalancing}. Since subscribers are shifted to new brokers, a number of updates have to be done in PRTs as a result of unsubscribe and resubscribe operations. In the event of broker or link failure, the topology repair process is complex due to the importance of maintaining the acyclic property of the overlay. Since brokers are aware of only their direct neighbours, the topology repair process is a challenging task: no broker has full knowledge of the overlay network, which can result in network partition \cite{reza-thesis}. 

Cyclic overlays can improve performance and throughput by offering multiple paths among publishers and subscribers. The best available path can be selected dynamically to route notifications, reducing delivery delays when a link congestion or failure is detected. Although multiple paths may be available from publishers to interested subscribers, selecting an optimal routing path is challenging. Further, detecting and eliminating message loops and extra messages generated in the \textit{Advertisement Broadcast Process (ABP)} consume valuable network resources inefficiently \cite{Li_ADAP}. Existing APS systems offer fixed end-to-end notification delivery paths in acyclic and cyclic overlays. Advertisements and matching subscriptions in cyclic overlays generate only a single path for notification forwarding. When a link is congested or broker fails, new routing paths have to be generated to avoid congested links or failed brokers in notifications routing. This requires eliminating effected path information in SRTs and PRTs and reissue advertisements and subscriptions calls to generate new routing paths to avoid congested links or overwhelmed brokers. This process is complex, and inefficient for large scale setups and is not suitable for delay sensitive applications. Content-based routing in cyclic overlays has received only a little attention and, existing PS systems offer a limited dynamic multi-path routing (cf. Section 9). 

This paper introduces \textit{OctopiA}, an APS system that offers efficient \textit{inter-cluster} dynamic routing of notifications. OctopiA uses \textit{Structured Cyclic Overlay Topology (SCOT)} that we introduced previously for content-based routing networks \cite{OctopiS}. OctopiA uses a clustering technique with a bit-vector mechanism to handle the issue of message loops in routing. The proposed advertisement forwarding algorithm exploits \textit{structuredness} of SCOT to forward advertisements to a selected part of the overlay and generate advertisement tree of length 1 (cf. Section 5). No extra message is generated in the advertisement forwarding process. OctopiA also introduces the concept of \textit{Subscription Subgrouping}, which divides issued subscriptions into exclusive subgroups of subscriptions. Each subgroup is broadcast to only a separate SCOT cluster, called \textit{Host Cluster} of the subgroup, while advertisements are used to connect publishers to interested subscribers present on different SCOT clusters (cf. Section 6). Since a subscription is broadcast to a separate cluster only (cf. Section 4), \textit{subscribe} or \textit{unsubscribe} calls from a subscriber generates updates in PRTs of brokers of subscriber's host cluster (also find typical computation word for covering and merging in unsub...). This paper makes the following contributions. 
\begin{enumerate}
	\item We introduce use of SCOT overlays to achieve inter-cluster dynamic routing in APS systems. Section 3 discusses how Cartesian Product of graphs can be used to formally describe a structured cyclic overlay. Constraints are introduced for the overlay to eliminate notification cycles (or loops).
	\item Section 4 describes a clustering approach and a lightweight cluster bit vector mechanism to eliminate cycles without padding path information in notifications. The bit-vector mechanism is used to achieve subscription subgrouping, which divides issued subscriptions into multiple disjoint subgroups where subscriptions in each subgroup is broadcast to a separate cluster. 
	\item A novel approach of advertisement forwarding in SCOT overlays is introduced in Section 5, while Section 6 discusses that how subscription broadcast is done in a cluster. The bit-vector mechanism that connects publishers with interested subscribers connected to brokers in different clusters is explained in Section 4.	 
	\item Section 7 presents details about static and dynamic notification routing algorithms implemented in OctopiA. We also implement advertisement tree identification based routing to compare state--of--the--art with our static and dynamic routing algorithms. 
\end{enumerate}
Background to the research problem and related work is discussed in Section 2. Section 8 evaluates and compares the routing algorithms presented in Section 7. The experimental results show that the static and dynamic routing algorithms for the cluster-based structured overlay perform better than identification-based routing algorithm for the unclustered structured cyclic overlay.  
\section{Background and Related Work}
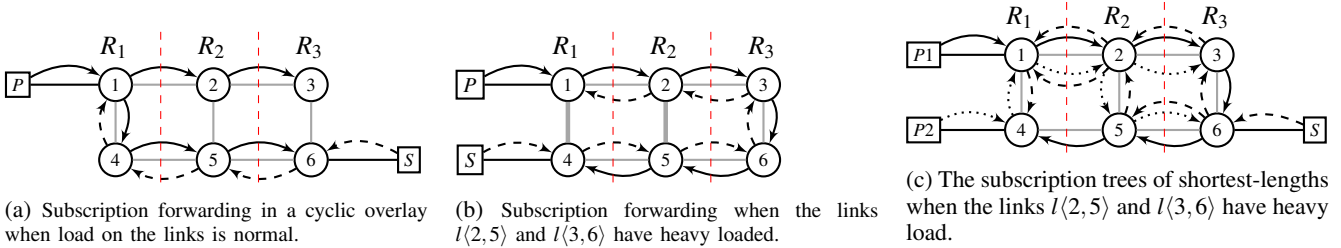
\begin{figure*}
	%define line/arrow styles...
	\tikzstyle{line} = [draw, -latex']
	\def\scaleFac {0.7}
	\begin{subfigure}[b]{0.31\textwidth}
		\begin{tikzpicture}
		\def\xInc {1.3}
		\def\x {0}
		\def\y {0}
		\def\yInc {1}
		\tikzstyle{every node} = [thick, scale=1]
		\node[draw, circle, scale=\scaleFac] (1) at (\x+\xInc,\y+\yInc) 		{$1$};
		\node[draw, circle, scale=\scaleFac] (2) at (\x+\xInc*2,\y+\yInc) 		{$2$};
		\node[draw, circle, scale=\scaleFac] (3) at (\x+\xInc*3,\y+\yInc) 		{$3$};
		\node[draw, circle, scale=\scaleFac] (4) at (\x+\xInc,\y) 				{$4$};
		\node[draw, circle, scale=\scaleFac] (5) at (\x+\xInc*2,\y) 			{$5$};
		\node[draw, circle, scale=\scaleFac] (6) at (\x+\xInc*3,\y) 			{$6$};
		%define the clients here...
		\node[draw, rectangle, scale=0.7] (S)	at (\x+\xInc*4,0) 						{$S$};
		\node[draw, rectangle, scale=0.7] (P)		 at (\x, \y+\yInc)		{$P$};
		
		\draw [color=gray!70, thick]  (5) -- (4) (6) -- (5) (3) -- (6) (2) -- (3) (1) -- (2) (4) -- (1) (5) -- (2); 
		\draw [thick] (S) -- (6) (P) -- (1);	
		
		%draw arrows for P
		\path [line, thick] (P) to [out=30,in=150] (1);
		\path [line, thick] (1) to [out=300,in=70] (4);
		\path [line, thick] (4) to [out=30,in=155] (5);
		\path [line, thick] (5) to [out=30,in=155] (6);
		\path [line, thick] (1) to [out=30,in=155] (2);
		\path [line, thick] (2) to [out=30,in=155] (3);
		%draw arrows for S
		\path [line, dashed, thick] (S) to [out=150,in=30] (6);
		\path [line, dashed, thick] (6) to [out=210,in=330] (5);
		\path [line, dashed, thick] (5) to [out=210,in=330] (4);	
		\path [line, dashed, thick] (4) to [out=120,in=240] (1);
		%draw the lines to indicate regions...
		\draw[color=red, dashed] (\x+1.9,-0.3) -- (\x+1.9,1.8);
		\draw[color=red, dashed] (\x+3.2,-0.3) -- (\x+3.2,1.8);		
		%draw the region text
		\node[text width=3cm] at (\x+\xInc*2, 1.5) {$R_1$};
		\node[text width=3cm] at (\x+\xInc*3, 1.5) {$R_2$};
		\node[text width=3cm] at (\x+\xInc*4, 1.5) {$R_3$};				
		\end{tikzpicture}
		\caption{\footnotesize Subscription forwarding in a cyclic overlay when load on the links is normal.}
		\label{fig:S1}
	\end{subfigure}
	~ %add desired spacing between images, e. g. ~, \quad, \qquad, \hfill etc.
	%(or a blank line to force the subfigure onto a new line)
	\begin{subfigure}[b]{0.31\textwidth}
		\begin{tikzpicture}
		\def\xInc {1.3}
		\def\x {0}
		\def\y {0}
		\def\yInc {1}
		\tikzstyle{every node} = [thick, scale=1]
		\node[draw, circle, scale=\scaleFac] (1) at (\x+\xInc,\y+\yInc) 		{$1$};
		\node[draw, circle, scale=\scaleFac] (2) at (\x+\xInc*2,\y+\yInc) 		{$2$};
		\node[draw, circle, scale=\scaleFac] (3) at (\x+\xInc*3,\y+\yInc) 		{$3$};
		\node[draw, circle, scale=\scaleFac] (4) at (\x+\xInc,\y) 				{$4$};
		\node[draw, circle, scale=\scaleFac] (5) at (\x+\xInc*2,\y) 			{$5$};
		\node[draw, circle, scale=\scaleFac] (6) at (\x+\xInc*3,\y) 			{$6$};
		%define the clients here...
		\node[draw, rectangle, scale=0.8] (S)	at (\x, 0) 				{$S$};
		\node[draw, rectangle, scale=0.8] (P)		 at (\x, \y+\yInc)			{$P$};
		
		\draw [color=gray!70, thick]  (5) -- (4) (6) -- (5) (3) -- (6) (2) -- (3) (1) -- (2); 
		\draw [thick] (S) -- (4) (P) -- (1);
		\draw [line width=0.06cm, gray!70] (4) -- (1) (5) -- (2);	
		%draw arrows for P
		\path [line, thick] (P) to [out=30,in=150] (1);
		\path [line, thick] (3) to [out=300,in=70] (6);
		\path [line, thick] (5) to [out=210,in=340] (4);
		\path [line, thick] (6) to [out=210,in=340] (5);
		\path [line, thick] (1) to [out=30,in=155] (2);
		\path [line, thick] (2) to [out=30,in=155] (3);
		%draw arrows for S
		\path [line, dashed, thick] (S) to [out=30,in=155] (4);
		\path [line, dashed, thick] (4) to [out=30,in=155] (5);
		\path [line, dashed, thick] (5) to [out=30,in=155] (6);	
		\path [line, dashed, thick] (6) to [out=120,in=240] (3);
		\path [line, dashed, thick] (3) to [out=210,in=340] (2);
		\path [line, dashed, thick] (2) to [out=210,in=340] (1);
		%draw the lines to indicate regions...
		\draw[color=red, dashed] (\x+1.9,-0.3) -- (\x+1.9,1.8);
		\draw[color=red, dashed] (\x+3.2,-0.3) -- (\x+3.2,1.8);		
		%draw the region text
		\node[text width=3cm] at (\x+\xInc*2, 1.5) {$R_1$};
		\node[text width=3cm] at (\x+\xInc*3, 1.5) {$R_2$};
		\node[text width=3cm] at (\x+\xInc*4, 1.5) {$R_3$};				
		\end{tikzpicture}
		\caption{\footnotesize Subscription forwarding when the links $l \langle 2,5\rangle$ and $l \langle 3,6\rangle$ have heavy loaded.}
		\label{fig:S2}
	\end{subfigure}
	~
	\begin{subfigure}[b]{0.31\textwidth}
		\begin{tikzpicture}
		\def\xInc {1.3}
		\def\x {0}
		\def\y {0}
		\def\yInc {1}
		\tikzstyle{every node} = [thick, scale=1]
		\node[draw, circle, scale=\scaleFac] (1) at (\x+\xInc,\y+\yInc) 		{$1$};
		\node[draw, circle, scale=\scaleFac] (2) at (\x+\xInc*2,\y+\yInc) 		{$2$};
		\node[draw, circle, scale=\scaleFac] (3) at (\x+\xInc*3,\y+\yInc) 		{$3$};
		\node[draw, circle, scale=\scaleFac] (4) at (\x+\xInc,\y) 				{$4$};
		\node[draw, circle, scale=\scaleFac] (5) at (\x+\xInc*2,\y) 			{$5$};
		\node[draw, circle, scale=\scaleFac] (6) at (\x+\xInc*3,\y) 			{$6$};
		%define the clients here...
		\node[draw, rectangle, scale=0.7] (P2)		 at (\x, 0) 				{$P2$};
		\node[draw, rectangle, scale=0.7] (P1)		 at (\x, \y+\yInc)			{$P1$};
		\node[draw, rectangle, scale=0.7] (S)		 at (\x+\xInc*4, 0)			{$S$};
		
		\draw [color=gray!70, thick]  (5) -- (4) (6) -- (5) (3) -- (6) (2) -- (3) (1) -- (2) (4) -- (1) (5) -- (2); 
		\draw [thick] (P2) -- (4) (P1) -- (1) (S) -- (6);	
		%draw arrows for P1
		\path [line, thick] (P1) to [out=30,in=150] (1);
		\path [line, thick] (3) to [out=300,in=70] (6);
		\path [line, thick] (5) to [out=210,in=340] (4);
		\path [line, thick] (6) to [out=210,in=340] (5);
		\path [line, thick] (1) to [out=30,in=155] (2);
		\path [line, thick] (2) to [out=30,in=155] (3);
		%draw arrows for P2
		\path [line, dotted, thick] (P2) to [out=30,in=155] (4);
		\path [line, dotted, thick] (4) to [out=120,in=250] (1);
		\path [line, dotted, thick] (1) to [out=330,in=210] (2);	
		\path [line, dotted, thick] (2) to [out=230,in=120] (5);
		\path [line, dotted, thick] (5) to [out=30,in=155] (6);
		\path [line, dotted, thick] (2) to [out=330,in=210] (3);
		%arrows for subscription tree of S...
		\path [line, dashed, thick] (S) to [out=155,in=25] (6);
		\path [line, dashed, thick] (6) to [out=120,in=240] (3);
		\path [line, dashed, thick] (6) to [out=135,in=45] (5);
		\path [line, dashed, thick] (3) to [out=135,in=45] (2);
		\path [line, dashed, thick] (2) to [out=135,in=45] (1);
		\path [line, dashed, thick] (5) to [out=70,in=290] (2);
		\path [line, dashed, thick] (2) to [out=230,in=310] (1);
		\path [line, dashed, thick] (1) to [out=290,in=70] (4);
		%draw the lines to indicate regions...
		\draw[color=red, dashed] (\x+1.9,-0.3) -- (\x+1.9,1.8);
		\draw[color=red, dashed] (\x+3.2,-0.3) -- (\x+3.2,1.8);		
		%draw the region text
		\node[text width=3cm] at (\x+\xInc*2, 1.5) {$R_1$};
		\node[text width=3cm] at (\x+\xInc*3, 1.5) {$R_2$};
		\node[text width=3cm] at (\x+\xInc*4, 1.5) {$R_3$};				
		\end{tikzpicture}
		\caption{The subscription trees of shortest-lengths when the links $l \langle 2,5\rangle$ and $l \langle 3,6\rangle$ have heavy load.}
		\label{fig:S3}
	\end{subfigure}
	\caption{\textit{A cyclic overlay of six brokers. The grey lines indicates links that connect the overlay brokers. The dashed, solid and dotted arrow messages indicate the subscription trees of S1, S2, and S3 respectively. The dashed red lines seprate regions.}} 
	\label{fig:Subscription}
\end{figure*}
This section describes the issue of cycles in APS systems. We also discuss that why, even with multiple paths between publishers and subscribers, dynamic routing of notification is a difficult task. 

Cycles generate redundant messages in content-based routing and it is increasingly important to detect and discard them. The motivation for advertisements in APS systems is to inform the event notification service about which kind of notifications will be generated by which publisher. The \textit{Advertisement Broadcast Process (ABP)} broadcasts an advertisement to network and forms a tree that reaches every broker. Each broker saves the advertisement in its SRT in the form of \{\textit{advertisement, lasthop}\} tuple where the lasthop indicates the last sender (publisher or broker) of the advertisement. An overlay broker propagates a subscription is reverse direction, along the trees of matching advertisements, thereby linking a publisher of interest with interested subscribers by activating those paths by saving the subscription in its PRT. Notifications are later on forwarded only onto activated paths. 
In cyclic overlays, duplicates of an advertisement may be received by a broker. An advertisement tree is assigned a unique identification to detect and discards duplicates \cite{Li_ADAP}. This process is further explained in Fig. 1, which shows a cyclic overlay of six brokers deployed across three geographical regions: $R_{1}$, $R_{2}$, and $R_{3}$. Brokers 1 and 4 are in $R_{1}$, brokers 2 and 5 are in $R_{2}$, and brokers 3 and 6 are in $R_{3}$. In this paper, an overlay link is represented as $l\langle source,destination\rangle$, where the \textit{source} identifies message sending broker and the \textit{destination} represents the message receiving broker. Each advertisement tree is uniquely identified by a tree identification \textit{(TID)} which is saved along the advertisement in SRT during the ABP. Each broker uses the TIDs to detect duplicates in the ABP and saves the incoming advertisement only if received for the first time. A subscriber bounds to TIDs of all matching advertisements to receive notifications from the publishers of interest. 

In Fig. 1(a), publisher P is connected to Broker 1, while an interested subscriber S, is connected to Broker 6. Broker 1, which is also root of the tree of the advertisement issued by P, assigns a unique TID to advertisement before forwarding it to other brokers. After receiving the advertisement from Broker 4, Broker 5 discard duplicate of the same advertisement when received from Broker 2. Similarly, Broker 6 discards the duplicate when received from Broker 3 \cite{Li_ADAP}. This indicates that despite using unique TID for each advertisement, extra inter-broker messages (IMs) still generate to detect duplicates. The advertisement tree shown in Fig. 1(a) is of the shortest length, which may also reduces lengths of the subscription trees. In this case, publications are processed by a fewer number of brokers in the routing paths, which improves throughput and reduces latency. 

Although the advertisement tree generated in Fig. 1(a) is of the shortest length, the ABP does not always generate shortest length advertisement trees. Length of an advertisement tree is effected by in-broker processing and inter-broker communication delays. This is shown in Fig. 1(b) where the length of the advertisement tree is twice of the length of the advertisement tree in Fig. 1(a). This is due to more load on the links $\langle 1,4 \rangle$ and $\langle 2,5 \rangle$ when P issued the advertisement. For example, Broker 5 receives the advertisement from Broker 6 and discards a duplicate received from Broker 2. Similarly, Broker 4 discards a duplicate received from Broker 1. The length of the subscription tree in this case is also larger than the length of the subscription tree in Fig. 1(a). Although P and S are in the same region, each notification is processed by brokers in the other two regions before it reaches Broker 4 (where the interested subscriber is hosted). Carzaniga \cite{carz_thesis} uses a latency-based distance-vector algorithm, with the RPF technique, to generate advertisement paths with minimal delay. However, the approach may generate advertisement trees of large length as described above. Lengths of advertisement trees can be updated by periodically generating advertisement calls. However, this approach would require frequent updates in SRT and as a result, updates in PRT, which is not suitable for large network setting. Ideally, an advertisement tree should always has the shortest length, even if some link have high loads when an advertisement is issued.  

Subscriptions cycles creates redundant IMs and cause publications route indefinitely among brokers with interested subscribers \cite{Li_ADAP}. To prevent cycles in the subscription forwarding process (SFP), each subscriber host broker binds the TID of the matching advertisements with the subscription and the subscription is forwarded along the paths set by the advertisements of bounded TIDs. A subscription may be bound to more than one TIDs depending on the number of advertisements that match with the subscription. Paths associated with the bounded TIDs are used in dynamic routing of publications provided their last hops are different. A broker may receive multiple copies of the same subscription associated with different TIDs and coming from different paths. For example in Fig. 1(c), subscription of S matches with advertisements of P1 and P2. Broker 6 receives the advertisement of P1 from Broker 3 and advertisement of P2 from Broker 5. Both advertisement trees share the links $\langle 1, 2\rangle$ and  $\langle 2, 3\rangle$. TIDs of advertisements of P1 and P2 are bound to subscription of S, which is received twice by Broker 2 from the last hops, Broker 3 and Broker 5. When the host broker of a publisher receives a notification, the notification is assigned TID of the advertisement tree of the publisher. Each broker in the routing paths use the TID to find matching subscriptions and forward the publications to the next destinations. This indicates that a notification has to carry TIDs bounds to the matching subscriptions \cite{Li_ADAP}.

Dynamic routing refers to the capability of a PS system to alter the routing path in response to overloaded or failed links and/or brokers. Content-based dynamic routing is only supported when there are alternative routes available between a publisher and its interested subscribers. Unfortunately, the ABP and SFP generates only one routing path using advertisements and matching subscriptions. Alternative routing paths can be generated (provided overlay links are available) dynamically when congestion or link/broker failures are detected, the process requires updates in SRTs and PRTs, which is not suitable for large scale overlay networks.
\section{Structured Cyclic Overlay}
Structured cyclic topologies have been used effectively in Parallel Computing and Product Networks (PN) for many years \cite{kemal1,kday1}. A PN is an interconnected network obtained by taking the product of two or more graphs. Many PNs have been proposed as popular topologies, including product tree, tori, hypercube, and butterfly, etc \cite{kemal1}. These topologies have regular structures that make a formal study of properties like symmetry, diameter, fault tolerance, degree, network partition, and parallel paths possible. Parallel paths can be used to expedite
parallel transfer of a large number of messages. Furthermore, they can be used as alternative routes when a node or a link failure occurs. A study shows that the repetition of patterns (or use of template) makes design and analysis of computer networks simple. More specifically, the use of graph
product is effective in describing the design templates that are used to define path redundancy and estimate fault tolerance in
computer networks beforehand \cite{theGeneralizedPUG}. We use the Cartesian Product of Undirected Graphs (CPUG) to design large, structured overlay networks based on small graph patterns. In addition, we introduce a clustering technique that divides a structured overlay into identifiable groups of brokers, in order to eliminate cycles without carrying a routing path identification with each notification.

\subsection {Cartesian Product of Undirected Graphs}
A graph is an ordered pair $G = (V_{G}, E_{G})$, where $V_{G}$ is a finite set of vertices and $E_{G}$ is set of edges or links that connect two vertices in $G$. The number of vertices of $G$ (called \textit{order}) is $\mid G\mid$ (or $|V_{G}|$). Similarly, the number of edges in $G$ is $\parallel G \parallel$ (or $|E_{G}|$). Two vertices $u, v \in V_{G}$ are adjacent or neighbours if the edge $uv \in E_{G}$. A graph in which each pair of vertices are connected by an edge is called a \textit{complete graph}. The path distance $d$ between two nodes $a,b$ in a graph is measured in the number of edges between them, and is given as $d \langle a,b \rangle$. The diameter of a graph G, represented as \textit{diam(G)}, is the shortest path between the two most distant nodes in $G$. The set of neighbours of a vertex $v \in G$, denoted as \textit{N(v)}, is called the \textit{degree} of $v$. The minimum degree of G, represented by $\delta(G)$, is given as: $\delta(G) := min\{N(v) \mid v \in V_{G}\}$. The maximum degree of G is denoted by $\Delta(G)$ is given as: $\Delta(G) := max\{N(v) \mid v \in V_{G}\}$. The minimum number of nodes whose deletion from $G$ disconnects it is called \textit{vertex connectivity} of $G$ and is denoted by $\kappa (G)$. Similarly, the minimum number of edges whose removal from $G$ disconnects $G$ is called \textit{edge connectivity} and is represented by $\lambda (G)$. The relations $\delta(G) \leqslant d(G) \leqslant \Delta(G)$ and $\kappa (G) \leqslant \lambda(G) \leqslant \delta(G)$ are well known in graph theory literature \cite{CPUG_Book}.

A graph product is a binary operation that takes two small graph operands---for example $G(V_{G}, E_{G})$ and $H(V_{H}, E_{H})$---to produces a large graph whose vertex set is given by $V_{G} \mathsf{X} V_{H}$. Many types of graph products exist, but the three fundamental types are the Cartesian product, the Direct product, and the Strong product \cite{CPUG_Book}. All three of these graph products have been investigated extensively in graph theory and used widely in PNs. The core concept behind these graphs products is the rule-based interconnection of vertices of the graph operands. The Cartesian product provides the most suitable link pattern for our research. The link pattern generated by Direct product makes routing complex, and although Strong product provides robust adjacencies (i.e., more edges) between vertices of the operand graphs, a high node degree exerts more load on an overlay broker. Furthermore, Strong product introduces a high level redundancy in overlay links, which makes message routing a difficult task. We discuss CPUG and its applications in the following. Details on the other two types of graph products are available in \cite{CPUG_Book}. 

The $CPUG$ of two graphs $G(V_{G}, E_{G})$ and $H(V_{H}, E_{H})$ is denoted by $G  \square  H$, with vertex set $V_{G \Box H}$ and set of edges $E_{G \Box H}$. Two vertices $(g, h) \in V_{G \Box H}$ and $(g', h') \in V_{G \Box H}$ are adjacent if $g=g'$ and $hh' \in E_{G \Box H}$ or $gg' \in E_{G \Box H}$ and $h=h'$. Formally, the sets of vertices and edges of a CPUG are given as.
\begin{equation}
V_{G \square H} = \{ (g, h) | g \in V_{G}  \wedge  h \in V_{H}\}
\end{equation}
\begin{equation}
\left.\begin{aligned}
E_{G \square H} = \{ \langle (g, h)(g', h') \rangle | (g=g', hh' \in E_{H}) \\ \vee (gg' \in E_{G}, h=h')\}
\end{aligned}
\right\}
\end{equation}
The operand graphs $G$ and $H$ are called factors of $G  \square  H$. CPUG is commutative---that is, $G  \Box  H = H  \square  G$. $G \square H$ is connected if both of the factors $G$ and $H$ are connected. The minimum degree of $G \square H$ is additive i.e. $\delta(G\square H) = \delta(G) + \delta(H)$ \cite{CPUG_Book}. The vertex connectivity $\kappa$ for $G \square H$ is given as (where $|V_{G}| \geq 2$, and $|V_{H}| \geq 2$):
\begin{equation}
\small
\kappa (G \square H) = min \{  \kappa(G) | H|, |G|\kappa(H), \delta(G) + \delta(H) \}
\end{equation}  
The edge connectivity $\lambda$ for $G \square H$ is defined as \cite{CPUG_Book}:
\begin{equation}
\small
\lambda (G \square H) = min \{  \lambda(G) | H|, |G|\lambda(H), \delta(G) + \delta(H) \}
\end{equation}
Although CPUG of $n$ number of graphs is possible, we are concerned with CPUG of only two graphs in this paper.
\subsection{Structured Cyclic Overlay Topology}
\textit{Structured Cyclic Overlay Topology (SCOT)}, represented as $\mathbb{S}$ in this paper, is a $CPUG$ of two graphs. One graph, represented by $G_{af}$, is called \textit{SCOT acyclic factor}, while the second graph operand, represented by $G_{cf}$, is called \textit{SCOT connectivity factor}. A SCOT has two important properties: (i) \textit{Acyclic Property} emphasizes that the $G_{af}$ must be an acyclic graph, and (ii) \textit{Connectivity Property} requires that $G_{cf}$ must be a complete graph. These properties augment a SCOT with essential characteristics that are used for generating routing paths of shortest lengths. $V_{af}$ and $V_{cf}$ are the sets of vertices of $G_{af}$ and $G_{cf}$, while $E_{af}$ and $E_{cf}$ are the sets of edges of $G_{af}$ and $G_{cf}$ respectively. For a \textit{singleton graph} of vertex set $\{h\} \subset V_{cf}$, the graph $G_{af}^h$ generated by $G_{af} \Box \{h\}$ is called a $G^h_{af}-fiber$ with \textit{index h}. Similarly, for a singleton graph of vertex set $\{m\} \subset G_{af}$, the graph $G^m_{cf}$ generated by $\{m\} \square G_{cf}$ is called a $G^m_{cf}-fiber$ with \textit{index m}. We describe the importance of using indexes in SCOT fibers in Section IV. The definitions of the fibers indicate that, for each vertex of $G_{cf}$, $CPUG$ generates one replica of $G_{af}$, and for each vertex of $G_{af}$, \textit{CPUG} generates one replica of $G_{cf}$. The number of distinct fibers of $G_{af}$ and $G_{cf}$ is equal to $|V_{cf}|$ and $|V_{af}|$ respectively. 
\begin{tikzpicture}
\tikzstyle{every node} = [minimum size=7mm]
\def\y {6}
\def\x {1}
\def\xInc {1}
\def\scaleFac {0.7}
\node[draw, thick, circle, scale=\scaleFac] (a) at (\x,\y) {$a$};
\node[draw, thick, circle, scale=\scaleFac] (b) at (\x+\xInc,\y) {$b$};
\node[draw, thick, circle, scale=\scaleFac] (c) at (\x+\xInc*2,\y) {$c$};
\node[draw, thick, circle, scale=\scaleFac] (d) at (\x+\xInc*3,\y) {$d$};
\node[draw, thick, circle, scale=\scaleFac] (e) at (\x+\xInc*4,\y) {$e$};
\node[draw, thick, circle, scale=\scaleFac] (f) at (\x+\xInc*5,\y) {$f$};
%draw the curved line...
\draw [thick] (b) to [out=40,in=140] (d);
\draw [thick] (a) -- (b) (b) -- (c) (d) -- (e) (e) -- (f);
%draw the operator box.
\node[draw, thick, rectangle, scale=0.4] (op)	at (\x+\xInc*6,\y) {};
%draw the connectivity triangle...
\node[draw, thick, circle, scale=\scaleFac, fill=green!50] (va) at (\x+\xInc*7,\y) {$1$};
\node[draw, thick, circle, scale=\scaleFac] (vb) at (\x+\xInc*8,\y) {$2$};
\node[draw, thick, circle, scale=\scaleFac, pattern=north west lines, pattern color=gray!40] (vc) at (\x+\xInc*7+0.5,\y+0.7) {$0$};		
%draw tirangle edges...
\draw [dashed, thick] (va) -- (vb) (vb) -- (vc) (vc) -- (va);
\end{tikzpicture}

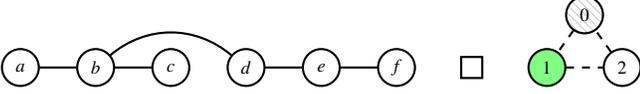
\captionof{figure}{\textit{Operands of CPUG: Left of $\Box$ is $G_{af}$ which is an H-graph; right of $\Box$ is $G_{cf}$ which is a triangle.}}
\label{fig:Operands}
\begin{tikzpicture}
%\footnotesize
\def\scaleSCOT {0.6}
\def\xInc {1.5}
\def\x {0}
\node[draw, thick, circle, scale=\scaleSCOT, pattern=north west lines, pattern color=gray!40] (1) at (\x,6) {$a,0$};
\node[draw, thick, circle, scale=\scaleSCOT, pattern=north west lines, pattern color=gray!40] (2) at (\x+\xInc,6) {$b,0$};
\node[draw, thick, circle, scale=\scaleSCOT, pattern=north west lines, pattern color=gray!40] (3) at (\x+\xInc*2,6) {$c,0$};
\node[draw, thick, circle, scale=\scaleSCOT, pattern=north west lines, pattern color=gray!40] (4) at (\x+\xInc*3,6) {$d,0$};
\node[draw, thick, circle, scale=\scaleSCOT, pattern=north west lines, pattern color=gray!40] (5) at (\x+\xInc*4,6) {$e,0$};
\node[draw, thick, circle, scale=\scaleSCOT, pattern=north west lines, pattern color=gray!40] (6) at (\x+\xInc*5,6) {$f,0$};
%draw the curved line...
\draw [thick] (2) to [out=30,in=150] (4);
\node[draw, thick, circle, scale=\scaleSCOT, fill=green!50] (11) at (\x,4.8) {$a,1$};
\node[draw, thick, circle, scale=\scaleSCOT, fill=green!50] (12) at (\x+\xInc,4.8) {$b,1$};
\node[draw, thick, circle, scale=\scaleSCOT, fill=green!50] (13) at (\x+\xInc*2,4.8) {$c,1$};
\node[draw, thick, circle, scale=\scaleSCOT, fill=green!50] (14) at (\x+\xInc*3,4.8) {$d,1$};
\node[draw, thick, circle, scale=\scaleSCOT, fill=green!50] (15) at (\x+\xInc*4,4.8) {$e,1$};
\node[draw, thick, circle, scale=\scaleSCOT, fill=green!50] (16) at (\x+\xInc*5,4.8) {$f,1$};
%draw the curved line...
\draw [thick] (12) to [out=30,in=150] (14);
\node[draw, thick, circle, scale=\scaleSCOT] (21) at (\x,3.6) {$a,2$};
\node[draw, thick, circle, scale=\scaleSCOT] (22) at (\x+\xInc,3.6) {$b,2$};
\node[draw, thick, circle, scale=\scaleSCOT] (23) at (\x+\xInc*2,3.6) {$c,2$};
\node[draw, thick, circle, scale=\scaleSCOT] (24) at (\x+\xInc*3,3.6) {$d,2$};
\node[draw, thick, circle, scale=\scaleSCOT] (25) at (\x+\xInc*4,3.6) {$e,2$};
\node[draw, thick, circle, scale=\scaleSCOT] (26) at (\x+\xInc*5,3.6) {$f,2$};
%draw the curved line...
\draw [thick] (22) to [out=30,in=150] (24);
\draw [thick] (1) -- (2) (2) -- (3) (4) -- (5) (5) -- (6) (11) -- (12) (12) -- (13) (14) -- (15) (15) -- (16) (21) -- (22) (22) -- (23) (24) -- (25) (25) -- (26);
%draw the dashed i-COL lines...					
\draw  [dashed, thick] (1) -- (11) (11) -- (21) (2) -- (12) (12) -- (22) (3) -- (13) (13) -- (23) (4) -- (14) (14) -- (24) 
(5) -- (15) (15) -- (25) (6) -- (16) (16) -- (26);
%drwa the curved dotted links...		
\draw [dashed, thick] (1) to [out=240,in=120] (21) (2) to [out=240,in=120] (22) (3) to [out=240,in=120] (23) (4) to [out=240,in=120] 
(24) (5) to [out=240,in=120] (25) (6) to [out=240,in=120] (26);
%draw cluster rectabgle 
\draw [blue,dotted,thick] (-0.5,6.7) -- (8,6.7) -- (8,5.5) -- (-0.5,5.5) -- (-0.5,6.7);
\draw [blue,dotted,thick] (-0.5,5.45) -- (8,5.45) -- (8,4.4) -- (-0.5,4.4) -- (-0.5,5.45);
\draw [blue,dotted,thick] (-0.5,4.35) -- (8,4.35) -- (8,3.2) -- (-0.5,3.2) -- (-0.5,4.35);
%draw the cluster label 
\node[text width=3cm] at (\x+9.1,6.4) {$C_0$};
\node[text width=3cm] at (\x+9.15,5.25) {$C_1$};
\node[text width=3cm] at (\x+9.15,4.1) {$C_2$};
\end{tikzpicture}

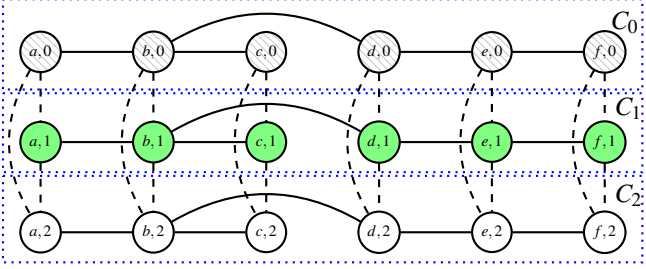
\captionof{figure}{\textit{Structured Cyclic Overlay Topology (SCOT) generated by CPUG operands shown in Fig. 2.}}
\label{fig:SCOT1}

In addition to acyclic and connectivity properties, a SCOT has two more properties: (i) \textit{Index property}, which emphasizes that the labels of nodes of $G_{cf}$ must be a sequence of unique integers starting from zero, and (ii) \textit{Label Order property}, which requires that the first operand (from left to right) of a $CPUG$ should be $G_{af}$. The index property implies that the index of each fiber of  $G_{af}$ is always an integer. The label order property indicates that the first part of the label of a SCOT node comes from the corresponding vertex of $G_{af}$, and the second part is the label of the corresponding vertex of $G_{cf}$, as indicated by Eq. 1. Reversing the order of operands does not generate extra links or nodes, since CPUG is commutative. In Fig. 2, the left operand of $\square$ operator is the $G_{af}$, while the second operand $G_{cf}$ is a triangle, which is a complete graph.\\
\textbf {Formal Representation ---}
A SCOT topology $\mathbb{S}$, with the set of vertices $V_{\mathbb{S}}$ and the set of edges $E_{\mathbb{S}}$, is formally represented using the \textit{Augmented Graph Union (AGU)} operator $\uplus$ in the following.
\begin{equation}
\mathbb{S} = \underset{i \in V_{cf}}{\biguplus} G^i_{af}
\label{eq:formal1}
\end{equation}
The AGU operator $\uplus$ combines vertices of all fibers of $G_{af}$ using traditional graph union pattern as $V_{\mathbb{S}} = \underset{i \in V_{cf}} \cup V^i_{af}$ and  $E_{\mathbb{S}}$ is expressed as.
\begin{equation}
E_{\mathbb{S}} = \Big( \underset{i \in V_{cf}} \bigcup E^i_{af}\Big) \bigcup \Big( \underset{j \in V_{af}} \bigcup E^j_{cf}\Big)
\label{eq:formal2}
\end{equation}
Equation 5 is the \textit{SCOT Acyclic Factor} (or \textit{SAF}) form of $\mathbb{S}$ while equation 6 is an alternative form of equation 2. The \textit{SCOT Connectivity Factor} (or \textit{SCF}) form of $\mathbb{S}$ is defined as.  
\begin{equation}
\mathbb{S} = \underset{j \in V_{af}}{\biguplus} G^j_{cf}
\label{eq:formal3}
\end{equation}
\section{Clustering}
In PS systems, clustering is the process of forming groups (i.e., clusters) of brokers dynamically in order to shrink routing paths to reduce delivery delays and IMs. \textit{Pure forwards} occur when a broker receives a notification but has no local subscribers interested in the notification. In this scenario, the broker works as the notification forwarder. Pure forwards occur frequently in large acyclic overlays when a publisher has a fewer interested subscribers connected to brokers in different overlay regions \cite{reza-thesis}. They cause inefficient use of network resources and are reduced by relocating subscribers. One advantage of dynamic clustering is the decrease in the number of pure forwards. Clustering in a SCOT is static, and clusters are identified before an overlay is deployed. Each cluster is assigned a unique identification, which is used in the routing of advertisements, subscriptions, and notifications. Each $G^i_{af}-fiber$ in a SCOT is treated as a separate group of brokers called a cluster, and is represented by $C_{i}$, where $i \in V_{cf}$. There are always $|V_{cf}|$ number of clusters in a SCOT. The SCOT in Fig. 3 contains three clusters (horizontal layers), each identified with a distinct label $C_{i}$, where $i \in \{0, 1, 2\}$ and is called the \textit{Cluster Index (CI)}. A CI is the label of a vertex of $G_{cf}$ that generates the cluster (or $G_{af}-fiber$) when a CPUG is calculated. Similarly, each $G^j_{cf}-fiber$ is called a \textit{Region} and is represented as $R_{j}$, where $j \in G_{af}$ and is known as the \textit{Region Index (RI)}. There are always $|V_{af}|$ number of regions in a SCOT. The SCOT in Fig. 3 contains six regions, each identified by a unique RI. The following are two valid relations and either can be used to formally represent $\mathbb{S}$.
\begin{equation}
\mathbb{S} = \underset{i \in V_{cf}}{\biguplus} C_{i}
\end{equation}

\begin{equation}
\mathbb{S} = \underset{j \in V_{af}}{\biguplus} R_{j}
\end{equation}
Equation 9 is the \textit{Cluster AGU (CAGU)} form of Equation \ref{eq:formal1}, while Equation 8 is the \textit{Region AGU (RAGU)} form of Equation \ref{eq:formal3}. This further shows that a SCOT overlay is an AGU of clusters and an AGU of regions as expressed by Equations9 and 8 respectively.

\subsection{Overlay Links and Messaging}
A SCOT has two types of links: (i) an \textit{intra-cluster overlay link (aLink)}, and (ii) an \textit{inter-cluster overlay link (iLink)}. aLinks connect brokers in the same cluster, while iLinks connect brokers in the same region. In this paper, solid lines and curves in each cluster in Fig. 3 are aLinks, connect brokers in each replica of $G_{af}$. The dotted lines and curves in each region in Fig. 3 are iLinks, which connect brokers in each replica of $G_{cf}$. Messaging along aLinks is referred to as \textit{intra-cluster messaging} while messaging along iLinks is referred to as \textit{inter-cluster messaging}. $\mathrm{L}^m_{a}$, the set of all aLinks in a cluster $C_{m}$ (where $m \in V_{cf}$) is given as: $\mathrm{L}^m_{a} = \{ l \langle (a, m,), (b, m) \rangle | a, b \in V_{af} \wedge ab \in E_{af} \wedge m \in V_{cf} \}$. The set of all aLinks in $\mathbb{S}$ is: $\underset{m \in V_{cf}}{\cup} \mathrm{L}^m_{a}$. Similarly, $\mathrm{L}^n_{i}$, the set of all iLinks in a region $R_{n}$ for an RI $n \in V_{af}$ is given as: $\mathrm{L}^n_{i} = \{ l \langle (n, u), (n, v) \rangle | n \in V_{af} \wedge u,v  \in V_{cf} \wedge uv \in E_{cf} \}$. The set of all iLinks in $\mathbb{S}$ is expressed as: $\mathrm{L}_{i} = \underset{j \in V_{af}}{\cup} \mathrm{L}^j_{i}$. There are $|E_{af}|.|V_{cf}|$ number of aLinks and $|V_{af}|.|E_{cf}|$ iLinks in $\mathbb{S}$, which shows that $E_{\mathbb{S}} = |E_{af}||V_{cf}| + |V_{af}||E_{cf}|$.
\subsection{Classification of Clusters and Brokers}
There are two types of clusters in a SCOT: (i) a \textit{Primary or Host Cluster (HC)}, and (ii) a \textit{Secondary Cluster (SC)}. The cluster that contains the host broker of a publisher is the host or primary cluster of the publisher, while all other clusters are the secondary clusters of the publisher. For example, in Fig. 5, $C_{0}$ is the host or primary cluster, while $C_{1}$, $C_{2}$ are secondary clusters of publisher P1 hosted by B(a, 0). 

The \textit{Primary neighbours} of a broker are its direct neighbours in the same cluster, while the \textit{Secondary neighbours} are those in the same region. No two brokers in a region belong to same cluster. This arrangement of brokers requires only one iLink for inter-cluster messaging. Clusters that host interested subscribers of a publisher are called its \textit{Target Clusters (TCs)}. An SC, which is a TC, is called \textit{Secondary Target Cluster (STC) } of the publisher. The host cluster of the publisher is its \textit{Primary Target Cluster (PTC)} if it hosts at least one subscriber interested in receiving notifications from the publisher. An \textit{edge broker} in a SCOT has at most one primary neighbour, while an \textit{inner broker} has at least two primary neighbours. The number of inner or edge brokers in a SCOT is the product of the number of inner or edge brokers in $G_{af}$ with $|V_{cf}|$. All brokers in a region are the same type (i.e., are either inner or edge). A SCOT broker is aware of its own type, the types of its primary and secondary neighbours, and the CI of its HC. In Fig. 5, $C_{0}$ (with CI 0) is the host cluster of P1 and P2. Host broker of P1 (i.e, B(a,0)) is an edge broker, while B(d, 0), which is the host broker of P2, is an inner broker. All brokers in the region $R_{a}$ are edge brokers, while  the region $R_{d}$, has only inner brokers.  

\subsection{Cluster Index Vector}
Cluster Index Vector (CIV) is a row (vector) of bits used to identify the host broker, and the STCs of a publisher. CIV is saved in SRTs with an advertisement and is used in advertisement forwarding (cf. Section 5), subscription broadcast (cf. Section 6) and dynamic routing of notifications (cf. Section 7). Bits in CIV are indexed from right to left, with the index of the right most bit being zero. Each bit of CIV, called the \textit{Cluster Index Bit (CIB)}, is reserved for a SCOT cluster where the index of the bit is same as the CI of the cluster it represents. CIV has three contexts: (i) \textit{the advertisement context ($A_{cxt}$)}, (ii) \textit{the subscription context ($S_{cxt}$)}, and (iii) \textit{the publication context ($P_{cxt}$)}. Each context has one \textit{Context Bit (CB)}, which bears special meaning to the context. Fig. 4 provides further details on the three contexts and their CBs. 

Fig. 4(a) shows CIV in $A_{cxt}$, which is saved with the advertisement at the host broker of a publisher $Pi$ that issued the advertisement. The CIV in $A_{cxt}$ is meaningful only at the host broker of $Pi$. The bit with black background is the CB of the advertisement and indicates the CI of the host cluster of $Pi$. All other bits with value 1 indicate the STCs of $Pi$. The CB in Fig. 4(a) indicates that $Pi$ is connected to a broker of $C_{2}$, and has two STCs: $C_{0}$ and $C_{1}$ (their bits in CIV are 1). The context of a CIV shifts from $A_{cxt}$ to $S_{cxt}$ when an advertisement is forwarded to secondary brokers of the host broker of the advertisement issuing publisher (cf. Section 5). Only two CIBs are meaningful in $S_{cxt}$ context: (i) CIB of the advertisement receiving SC (also called the \textit{Secondary Cluster Bit (SCB)} in CIV in $S_{cxt}$ context), (ii) CIB of the advertisement sending cluster. Fig. 4(b) shows that the advertisement is issued from a broker of $C_{2}$ while CI of the advertisement receiving cluster is 1. When the value of the SCB is 0, no subscription saved in PRT of brokers in the advertisement receiving cluster matches with the advertisement. The SCB is set to 1 if a matching subscription is saved in PRT of the advertisement receiving cluster (more details in Section 5). The final context of a CIV is the $P_{cxt}$, which is used in dynamic routing of notifications. The CIBs with values 1 indicate STCs of the publisher that issued the advertisement (cf. Section 7). The CIV in $P_{cxt}$ context shown in Fig. 4(c) indicates that the publisher is hosted by a broker in $C_{2}$ and $C_{0}$, and $C_{1}$ are STCs of the publisher (also talk about which context is saved in routing table and which is not saved in a routing table). 

Constructing an underlay--aware cyclic topology for a PS system deployed across different geographic regions is difficult. A SCOT clusters can be deployed in different geographic regions and a cluster's brokers can be placed and connected by respecting some conformity with the local underlay network topology. However, achieving the desired level of conformity is difficult, as all SCOT clusters have the same acyclic topology. A method to relax this restriction for topology structure will be part of the future work.
%define line/arrow styles...
\begin{tikzpicture}
\def\y {0.25}
% draw boxes
\foreach \x in {-0.5, 0, 0.5, 1, 1.5}
\node[draw, rectangle, scale=2.1, thick] (B1)	at  (\x, 1) {};
% draw index
\node at  (-0.5, 1.5) {4};
\node at  (0, 1.5) {3};
\node at  (0.5, 1.5) {2};
\node at  (1, 1.5) {1};
\node at  (1.5, 1.5) {0};
% text for bits
\foreach \x in {-0.5, 0}
\node at  (\x, 1) {0};

\node[fill=black, text=white] at  (0.5, 1) {1};
\node at  (1, 1) {1};
\node at  (1.5, 1) {1};

\def\a {3}
\draw [decorate,decoration={brace,amplitude=4pt}, rotate=0, thick] (-0.5, 1.7) -- (1.5, 1.7) node [black,midway,yshift=0.4cm] {\textit{Bit Indexes}};
\draw [decoration={brace,mirror,raise=5pt, amplitude=4pt},decorate, thick] (-0.7, 0.85) --  node[below=10pt]{\textit{Bit Values}} (1.7, 0.85) ; 
%start of the second grid ...
\foreach \x in {-0.5, 0, 0.5, 1, 1.5}
\node[draw, rectangle, scale=2.1, thick] (B2)	at  (\x+\a, 1) {};
% draw index ...
\node at  (-0.5+\a, 1.5) {4};
\node at  (0+\a, 1.5) {3};
\node at  (0.5+\a, 1.5) {2};
\node at  (1+\a, 1.5) {1};
\node at  (1.5+\a, 1.5) {0};

\foreach \x in {-0.5, 0, 1.5}
\node at  (\x+\a, 1) {0};

\node [pattern=north west lines, pattern color=gray!60] at  (1+\a, 1) {1};
\node [fill=black, text=white] at (0.5+\a, 1) {1};

\draw [decorate,decoration={brace,amplitude=4pt}, rotate=0, thick] (-0.5 + \a, 1.7) -- (1.5 + \a, 1.7) node [black,midway,yshift=0.4cm] {\textit{Bit Indexes}};
\draw [decoration={brace,mirror,raise=5pt, amplitude=4pt},decorate, thick] (-0.7+\a, 0.85) --  node[below=10pt]{\textit{Bit Values}}(1.7+\a, 0.85); 

\def\b {6}
\draw [decorate,decoration={brace,amplitude=4pt}, rotate=0, thick] (-0.5, 1.7) -- (1.5, 1.7) node [black,midway,yshift=0.4cm] {\textit{Bit Indexes}};
\draw [decoration={brace,mirror,raise=5pt, amplitude=4pt},decorate, thick] (-0.7, 0.85) --  node[below=10pt]{\textit{Bit Values}} (1.7, 0.85) ; 
%start of the second grid ...
\foreach \x in {-0.5, 0, 0.5, 1, 1.5}
\node[draw, rectangle, scale=2.1, thick] (B2)	at  (\x+\b, 1) {};
% draw index ...
\node at  (-0.5+\b, 1.5) {4};
\node at  (0+\b, 1.5) {3};
\node at  (0.5+\b, 1.5) {2};
\node at  (1+\b, 1.5) {1};
\node at  (1.5+\b, 1.5) {0};

\foreach \x in {0, 1, 1.5}
\node at  (\x+\b, 1) {1};

\node[fill=black, text=white] at  (0.5+\b, 1) {1};
\node at  (-0.5+\b, 1) {0};

\draw [decorate,decoration={brace,amplitude=4pt}, rotate=0, thick] (-0.5 + \b, 1.7) -- (1.5 + \b, 1.7) node [black,midway,yshift=0.4cm] {\textit{Bit Indexes}};
\draw [decoration={brace,mirror,raise=5pt, amplitude=4pt},decorate, thick] (-0.7+\b, 0.85) --  node[below=10pt]{\textit{Bit Values}}(1.7+\b, 0.85); 
\end{tikzpicture}

\begin{tikzpicture}		
\node [text width=3cm] at (-0.7, 0)  {(a) $A_{cxt}$ context};
\node [text width=3cm] at (1.8, 0)  {(b) $S_{cxt}$ context};
\node [text width=3cm] at (4.8, 0)  {(c) $P_{cxt}$ context};
\end{tikzpicture}

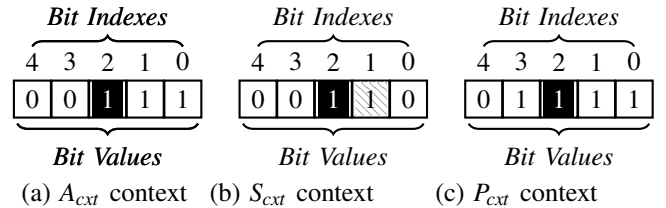
\captionof {figure}{\textit{(Left) CBV of a subscription indicates that the subscriber is hosted by a broker in cluster $C_{2}$. (Right) CBV of a notification indicates that $C_{0}$ and $C_{2}$ are the secondary target clusters.}}	
\label{fig:SRA}			
\section{Advertisement Forwarding}
In traditional APS systems, the ABP broadcasts an advertisement to all brokers in an overlay to form advertisement tree and any broker in an overlay network can match a subscription. A broker matches saved advertisements with received subscription to generate routing paths for forwarding notifications to interested subscribers \cite{SIENA_WIDE_AREA}.

OctopusA uses a different approach and broadcasts an advertisement only to brokers in the region of the host broker of a publisher. This implies that an advertisement is available only on the $|V_{cf}|$ number of brokers in SCOT and only at one broker in each cluster. This pattern of advertisements forwarding has particular benefits: size of SRTs reduced dramatically; updates are generated only in one region when a publisher advertises or unadvertises an advertisement. Due to connectivity property of the $G_{cf}$, size of each advertisement tree is always 1. The advertisement broadcast process in a SCOT is further explained in Fig. 5, where host cluster of the publisher P1 and P2 is $C_{0}$ and host clusters of the publishers P3 and P4 are $C_{1}$, and $C_{2}$ respectively. When a broker receives an advertisement from a publisher, it creates CIV with $A_{cxt}$ context, sets the CIB of the host cluster ON and all other bits OFF and saves a copy of the advertisement and CIV in SRT of the publisher's host cluster. In next step, the advertisement is forwarded to all secondary brokers of the publisher's host broker. Upon receiving the advertisement, each secondary broker performs two tasks: (i) turns the SCB ON if one or more subscriptions saved in its PRT match with the received advertisement, (ii) saves the received advertisement and CIV in SRT. If SCB is turned ON, the advertisement receiving secondary broker sends a $CIB\_SET$ message to the advertisement sending broker, which is also host of the advertisement issuing publisher. The $CIB\_SET$ message receiving broker sets the CIB of the message sending cluster ON in CIV of the advertisement. For each advertisement, at most one $CIB\_SET$ message is sent even if multiple interested subscribers are hosted by brokers of an STC (cf. Section 6). In the proposed AFP in SCOT, no identification is assigned to advertisement trees and no extra message are generated to detect and discard duplicates.

The advertisement forwarding process is further explained in the algorithm 1 using the scotAFP calls...

\begin{figure}
	\begin{tikzpicture}
	\def\scaleSCOT {0.7}
	\def\scaleBox {0.7}
	%define line/arrow styles...
	\tikzstyle{line} = [draw, -latex']
	\tikzstyle{lineR} = [draw, latex-']
	\def\y {6}
	\def\x {0}
	\def\xInc {1.2}
	\def\yInc {1.2}		
	\node[draw, circle, scale=\scaleSCOT, pattern=north west lines, pattern color=gray!40] (1) at (\x,\y) 			{$a,0$};
	\node[draw, circle, scale=\scaleSCOT, pattern=north west lines, pattern color=gray!40] (2) at (\x + \xInc*1,\y) {$b,0$};
	\node[draw, circle, scale=\scaleSCOT, pattern=north west lines, pattern color=gray!40] (3) at (\x+\xInc*2,\y)	{$c,0$};
	\node[draw, circle, scale=\scaleSCOT, pattern=north west lines, pattern color=gray!40] (4) at (\x+\xInc*3,\y) 	{$d,0$};
	\node[draw, circle, scale=\scaleSCOT, pattern=north west lines, pattern color=gray!40] (5) at (\x+\xInc*4,\y) 	{$e,0$};
	\node[draw, circle, scale=\scaleSCOT, pattern=north west lines, pattern color=gray!40] (6) at (\x+\xInc*5,\y) 	{$f,0$};
	\node[draw, circle, scale=\scaleSCOT, fill=green!50] (11) at (\x,\y-\yInc) 			{$a,1$};
	\node[draw, circle, scale=\scaleSCOT, fill=green!50] (12) at (\x+ \xInc*1,\y-\yInc) {$b,1$};
	\node[draw, circle, scale=\scaleSCOT, fill=green!50] (13) at (\x+ \xInc*2,\y-\yInc) {$c,1$};
	\node[draw, circle, scale=\scaleSCOT, fill=green!50] (14) at (\x+ \xInc*3,\y-\yInc) {$d,1$};
	\node[draw, circle, scale=\scaleSCOT, fill=green!50] (15) at (\x+ \xInc*4,\y-\yInc) {$e,1$};
	\node[draw, circle, scale=\scaleSCOT, fill=green!50] (16) at (\x+ \xInc*5,\y-\yInc) {$f,1$};
	%draw brokers...
	\node[draw, circle, scale=\scaleSCOT] (21) at (\x,\y-\yInc*2) 		   {$a,2$};
	\node[draw, circle, scale=\scaleSCOT] (22) at (\x+ \xInc*1,\y-\yInc*2) {$b,2$};
	\node[draw, circle, scale=\scaleSCOT] (23) at (\x+ \xInc*2,\y-\yInc*2) {$c,2$};
	\node[draw, circle, scale=\scaleSCOT] (24) at (\x+ \xInc*3,\y-\yInc*2) {$d,2$};
	\node[draw, circle, scale=\scaleSCOT] (25) at (\x+ \xInc*4,\y-\yInc*2) {$e,2$};
	\node[draw, circle, scale=\scaleSCOT] (26) at (\x+ \xInc*5,\y-\yInc*2) {$f,2$};
	%draw both the clients S2 and S3
	\node[draw, rectangle, scale=\scaleBox, text=red] (P1)	at  (\x,\y+\yInc-0.2) {$P1$};
	\node[draw, rectangle, scale=\scaleBox, text=red] (P2)	at (\x+\xInc*3,\y+\yInc-0.2) {$P2$};
	\node[draw, rectangle, scale=\scaleBox, text=blue] (P3)	at (\x+ \xInc*5+1, \y-\yInc) {$P3$};
	\node[draw, rectangle, scale=\scaleBox] (P4) at  (\x-1,\y-\yInc*2) {$P4$};
	%draw the curved line...
	\draw [thick, gray!70] (22) to [out=25,in=155] (24) (12) to [out=25,in=155] (14) (2) to [out=25,in=155] (4);
	\draw [thick, gray!70] (1) -- (2) (2) -- (3) (4) -- (5) (5) -- (6) (11) -- (12) (12) -- (13) (14) -- (15) (15) -- (16) (21) -- (22) (22) -- (23) (24) -- (25) (25) -- (26);
	\draw  (P1) -- (1) (P2) -- (4) (P3) -- (16) (P4) -- (21); 
	%draw the dashed i-COL lines...					
	\draw  [dashed, thick, gray!70] (1) -- (11) (11) -- (21) (2) -- (12) (12) -- (22) (3) -- (13) (13) -- (23) (4) -- (14) (14) -- (24) 
	(5) -- (15) (15) -- (25) (6) -- (16) (16) -- (26);
	%draw the curved dotted links...		
	\draw [dashed, thick, gray!70] (1) to [out=240,in=120] (21) (2) to [out=240,in=120] (22) (3) to [out=240,in=120] (23) (4) to [out=240,in=120] 
	(24) (5) to [out=240,in=120] (25) (6) to [out=240,in=120] (26);
	%draw the tree links of P1.
	\draw [line, thick, dashed] (P1) to [out=240,in=110] (1);
	\draw [line, thick, dashed] (1) to [out=240,in=110] (11);
	\draw [line, thick, dashed] (1) to [out=230,in=130] (21);
	%draw the tree link of P2.
	\draw [line, thick, dotted] (P2) to [out=240,in=110] (4);
	\draw [line, thick, dotted] (4) to [out=240,in=110] (14); 
	\draw [line, thick, dotted] (4) to [out=230,in=130] (24);
	%draw the tree link of P3.
	\draw [line, thick, blue] (P3) to [out=210,in=340] (16);
	\draw [line, thick, blue] (16) to [out=60,in=310] (6); 
	\draw [line, thick, blue] (16) to [out=310,in=60] (26);
	%draw the tree links of P4.
	\draw [line, thick] (P4) to [out=30,in=150] (21);
	\draw [line, thick] (21) to [out=60,in=300] (11);
	\draw [line, thick] (21) to [out=60,in=300] (1);
	\end{tikzpicture}
	\captionof{figure}{\small \textit{The two-step subscription forwarding process in the SCOT shown in Figure 3. The solid arrows indicate part of the subscription tree generated in the first step and the dashed arrows indicate part of the subscription tree generated in the second step.}}
	\label{fig:theAFP}
\end{figure}
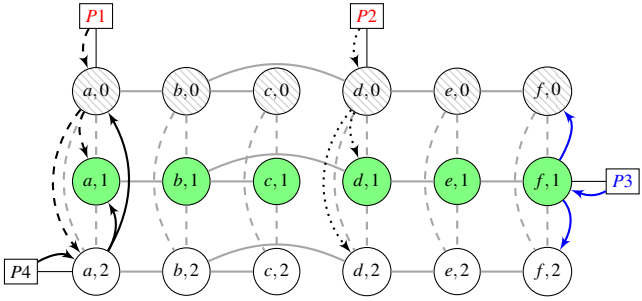

\begin {algorithm}
\small
\KwIn{$a:$ an advertisement message\;}
\KwOut{$DL:$ a list of next destinations that should receive $a$ \;}
\If{$isHostBroker (a)$}{
	$a.CIV \gets setBitOn(CI)$\;
	$a.hostBroker \gets false$\;
	%send this sub message to all primary neighbours
	\ForEach {$n \in SecondaryNeighbours$}{
		$a.next \gets n$\;
		$DL.add(a)$\;		
	}
	$SRT.add(a)$\;
} \ElseIf {$a.type = CIB\_SET$}{
$SRT.setCIB(a)$ \;
}\Else{
\If{$isMatchingSub(a) = true$}{
	$a.type \gets CIB\_SET$\;
	$a.CIV \gets setCIB(CI, a.CIV)$\;
	$scotAFP(a)$\;
}
$SRT.add(a)$\;
}
\BlankLine		
\label{algo:AFP}
\caption{$scotAFP(a)$}
\end{algorithm}

\section {Subscription Broadcast}
In traditional APS systems, advertisements are broadcast to an overlay network and matching subscriptions are used to create notification routing paths. This enables use of the RPF technique to route notifications only to those brokers that host interested subscribers \cite{SIENA_WIDE_AREA}. In OctopiA, each advertisement is available at most one broker of a cluster because an advertisement forwarded to only secondary neighbours of the host broker of the advertisement issuing publisher and, therefore, use of the RPF is not possible. 

A subscription in proposed approach is broadcast to brokers in the subscriber's host cluster. As each cluster, being a replica of $G_{af}$, is an acyclic overlay, no loops appear in subscription broadcast because no subscription is forwarded to any secondary brokers. Broadcast of a subscription in the host cluster of the subscriber is essential to find matching advertisements because each matching advertisement is available at only one unknown broker in the subscriber's host cluster. Subscription broadcast in host cluster is prerequisite to achieve inter-cluster dynamic routing (cf. Section 7). SBP is further explained in Fig. 6, where the host cluster of subscribers S1 and S2 is $C_{0}$, while the host cluster of subscribers S3 and S4 are $C_{1}$ and $C_{2}$ respectively. Subscription trees are generated according to approach described in \cite{SIENA_WIDE_AREA} for acyclic overlays. The link $l \langle(b,0),(c,0)\rangle$ in $C_{0}$ is part of the subscription trees of S1 and S2. Each broker, after receiving a subscription, performs two tasks: (i) saves the subscription in PRT, and (ii) sends \textit{CIB\_SET} message to those secondary brokers which host publishers with matching advertisements and their SCBs in CIVs of the matching advertisements are OFF. 

The $CIB\_SET$ message is sent only when an SCB in CIV is OFF. The purpose of sending $CIB\_SET$ message to the host broker of a publisher with matching advertisement is to set CIB of an SC ON because the SC has subscribers interested in receiving notifications from the publisher that issued the matching advertisement. If SCB is already ON, no $CIB\_SET$ message is sent to the host broker of the publisher. Host broker of a publisher forwards a notification to brokers of those STCs which have their CIBs 1 in the CIV in $A_{cxt}$ context. 

An SCB is used to stop sending multiple \textit{CIB\_SET} messages to host broker of a publisher that issued the advertisement. An SCB with 1 indicates that the host broker of the publisher is aware of forwarding a notification so no \textit{CIB\_SET} message should be sent. This is important to mention that each \textit{CIB\_SET} messages also carries the new matching subscription that generated the \textit{CIB\_SET} message. The \textit{CIB\_SET} message receiving broker finds the matching advertisement using the received subscription and sets the CIB of the message sending cluster ON (if it is previously OFF). Subscriptions received with the \textit{CIB\_SET} messages are not saved in PRT of the host broker of the publisher and notifications are routed to STCs using their CIBs in CIV in $A_{cxt}$ context. The maximum length of a subscription tree generated using technique is $diam(G_{af})$.

The SFP introduced in this section divides issued subscriptions into disjoint subgroups where each subgroup is broadcast to a separate cluster. This phenomenon is called \textit{Subscription Subgrouping} which uses CIVs in $A_{cxt}$ context to forward notifications to STCs. Subscription Subgrouping has multiple advantages. Updates caused by the use of covering and merging techniques when a subscriber subscribers or unsubscribes methods are called do not disrupt network traffic in other clusters. This shows that high churn rate on a cluster causes traffic load on a limited part of an overlay network. As subscriptions are limited to a SCOT cluster, interruption due to an overlay link or broker failure is limited to a cluster. Different notification access and QoS policies can be enforced on each cluster (qos-survey-paper). Subscription broadcast at cluster level has substantial benefits in dynamic inter-cluster routing of notifications (cf. Section ?).   

\begin{figure}
	\begin{tikzpicture}
	\def\scaleSCOT {0.7}
	\def\scaleBox {0.8}
	%define line/arrow styles...
	\tikzstyle{line} = [draw, -latex']
	\tikzstyle{lineR} = [draw, latex-']
	\def\y {6}
	\def\x {0}
	\def\xInc {1.2}
	\def\yInc {1.2}		
	\node[draw, circle, scale=\scaleSCOT, pattern=north west lines, pattern color=gray!40] (1) at (\x,\y) 			{$a,0$};
	\node[draw, circle, scale=\scaleSCOT, pattern=north west lines, pattern color=gray!40] (2) at (\x + \xInc*1,\y) {$b,0$};
	\node[draw, circle, scale=\scaleSCOT, pattern=north west lines, pattern color=gray!40] (3) at (\x+\xInc*2,\y)	{$c,0$};
	\node[draw, circle, scale=\scaleSCOT, pattern=north west lines, pattern color=gray!40] (4) at (\x+\xInc*3,\y) 	{$d,0$};
	\node[draw, circle, scale=\scaleSCOT, pattern=north west lines, pattern color=gray!40] (5) at (\x+\xInc*4,\y) 	{$e,0$};
	\node[draw, circle, scale=\scaleSCOT, pattern=north west lines, pattern color=gray!40] (6) at (\x+\xInc*5,\y) 	{$f,0$};
	\node[draw, circle, scale=\scaleSCOT, fill=green!50] (11) at (\x,\y-\yInc) 			{$a,1$};
	\node[draw, circle, scale=\scaleSCOT, fill=green!50] (12) at (\x+ \xInc*1,\y-\yInc) {$b,1$};
	\node[draw, circle, scale=\scaleSCOT, fill=green!50] (13) at (\x+ \xInc*2,\y-\yInc) {$c,1$};
	\node[draw, circle, scale=\scaleSCOT, fill=green!50] (14) at (\x+ \xInc*3,\y-\yInc) {$d,1$};
	\node[draw, circle, scale=\scaleSCOT, fill=green!50] (15) at (\x+ \xInc*4,\y-\yInc) {$e,1$};
	\node[draw, circle, scale=\scaleSCOT, fill=green!50] (16) at (\x+ \xInc*5,\y-\yInc) {$f,1$};
	%draw brokers...
	\node[draw, circle, scale=\scaleSCOT] (21) at (\x,\y-\yInc*2) 		   {$a,2$};
	\node[draw, circle, scale=\scaleSCOT] (22) at (\x+ \xInc*1,\y-\yInc*2) {$b,2$};
	\node[draw, circle, scale=\scaleSCOT] (23) at (\x+ \xInc*2,\y-\yInc*2) {$c,2$};
	\node[draw, circle, scale=\scaleSCOT] (24) at (\x+ \xInc*3,\y-\yInc*2) {$d,2$};
	\node[draw, circle, scale=\scaleSCOT] (25) at (\x+ \xInc*4,\y-\yInc*2) {$e,2$};
	\node[draw, circle, scale=\scaleSCOT] (26) at (\x+ \xInc*5,\y-\yInc*2) {$f,2$};
	%draw both the clients S2 and S3
	\node[draw, rectangle, scale=\scaleBox, text=red] (S1)	at  (\x-1,\y) 					{$S1$};
	\node[draw, rectangle, scale=\scaleBox, text=red] (S2)	at (\x+ \xInc*5+1, \y) 			{$S2$};
	\node[draw, rectangle, scale=\scaleBox, text=blue] (S3)	at (\x+ \xInc*5+1, \y-\yInc) 	{$S3$};
	\node[draw, rectangle, scale=\scaleBox] (S4) at  (\x-1,\y-\yInc*2) 						{$S4$};
	%draw the curved line...
	\draw [thick, gray!70] (22) to [out=25,in=155] (24) (12) to [out=25,in=155] (14) (2) to [out=25,in=155] (4);
	\draw [thick, gray!70] (1) -- (2) (2) -- (3) (4) -- (5) (5) -- (6) (11) -- (12) (12) -- (13) (14) -- (15) (15) -- (16) (21) -- (22) (22) -- (23) (24) -- (25) (25) -- (26);
	%draw the dashed i-COL lines...					
	\draw  [dashed, thick, gray!70] (1) -- (11) (11) -- (21) (2) -- (12) (12) -- (22) (3) -- (13) (13) -- (23) (4) -- (14) (14) -- (24) 
	(5) -- (15) (15) -- (25) (6) -- (16) (16) -- (26);
	%draw the curved dotted links...		
	\draw [dashed, thick, gray!70] (1) to [out=240,in=120] (21) (2) to [out=240,in=120] (22) (3) to [out=240,in=120] (23) (4) to [out=240,in=120] 
	(24) (5) to [out=240,in=120] (25) (6) to [out=240,in=120] (26);
	%draw the tree links of S1.
	\draw [line, thick, dashed] (S1) to [out=15,in=165] (1);
	\draw [line, thick, dashed] (1) to [out=15,in=165] (2);
	\draw [line, thick, dashed] (2) to [out=15,in=165] (3);
	\draw [line, thick, dashed] (5) to [out=15,in=165] (6);
	\draw [line, thick, dashed] (4) to [out=15,in=165] (5);
	\draw [line, thick, dashed] (2) to [out=30,in=150] (4);
	%draw the tree link of S2.
	\draw [line, thick, dotted] (S2) to [out=210,in=340] (6);
	\draw [line, thick, dotted] (2) to [out=210,in=330] (1); 
	\draw [line, thick, dotted] (6) to [out=210,in=340] (5);
	\draw [line, thick, dotted] (5) to [out=210,in=340] (4); 
	\draw [line, thick, dotted] (4) to [out=210,in=330] (2);
	\draw [line, thick, dotted] (2) to [out=330,in=210] (3);
	%draw the tree link of S3.
	\draw [line, thick, blue] (S3) to [out=210,in=340] (16);
	\draw [line, thick, blue] (12) to [out=210,in=330] (11); 
	\draw [line, thick, blue] (16) to [out=210,in=340] (15);
	\draw [line, thick, blue] (15) to [out=210,in=340] (14); 
	\draw [line, thick, blue] (14) to [out=210,in=330] (12);
	\draw [line, thick, blue] (12) to [out=330,in=210] (13);
	%draw the tree links of S4.
	\draw [line, thick] (S4) to [out=15,in=165] (21);
	\draw [line, thick] (21) to [out=15,in=165] (22);
	\draw [line, thick] (22) to [out=15,in=165] (23);
	\draw [line, thick] (25) to [out=15,in=165] (26);
	\draw [line, thick] (24) to [out=15,in=165] (25);
	\draw [line, thick] (22) to [out=30,in=150] (24);
	
	\draw  (S1) -- (1) (S2) -- (6); 
	\draw  (S3) -- (16); 
	\draw (S4) -- (21); 
	\end{tikzpicture}
	\captionof{figure}{\small \textit{The two-step subscription forwarding process in the SCOT shown in Figure 3. The solid arrows indicate part of the subscription tree generated in the first step and the dashed arrows indicate part of the subscription tree generated in the second step. }}
	\label{fig:theSFP}
\end{figure}
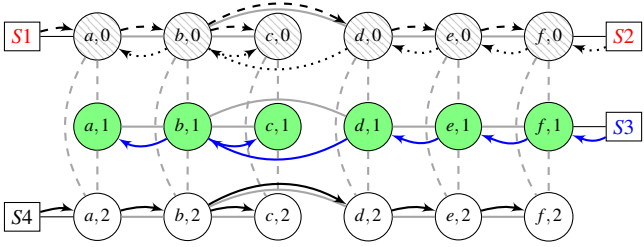

\begin {algorithm}
\small
\KwIn{$s:$ a subscription message\;}
\KwOut{$DL:$ a list of next destinations that should receive $s$ \;}

\ForEach {$n \in (PrimaryNeighbours - Sender)$}{
	$s.next \gets n$\;
	$DL.add(s)$\;
}	

\textit{/* Get only one matching advertisement per secondary cluster. Selected advertisements should have CB OFF. */}\\
$advs \gets getUniqueMatchingAdvs(s)$\;

\If{$advs.size > 0$}{
	\ForEach {$a \in advs$}{
		\If{$isCbON(a.CIV) \neq true$}{
			$CIB\_SET.next \gets a.lastHop$\;
			$CIB\_SET.sub \gets s$\;
			$DL.add(CIB\_SET)$\;
		}
	}
}
\label{algo:A1}
\caption{$scotSFP(s)$}
\end{algorithm}
\section{Notification Routing}
\begin{figure*}
	\tikzstyle{line} = [draw, -latex']
	\def\scaleFac {0.7}
	\begin{subfigure}[b]{0.23\textwidth}
		\begin{tikzpicture}
		\tikzstyle{line} = [draw, -latex']
		\centering
		\def\scaleSCOT {0.6}
		\def\scaleBox {0.7}
		\def\x {-0.5} 
		\def\xInc {1.4}
		\def\y {0} 
		\def\yInc {0.9}
		\def\xIncFirst {0.8}
		%begin the first topology diagram here ...
		\tikzstyle{every node} = [thick]
		\node[draw, circle, scale=\scaleSCOT] (1) at (\x,\y+\yInc*2) {$a,0$};
		\node[draw, circle, scale=\scaleSCOT] (2) at (\x+\xInc,\y+\yInc*2) {$b,0$};
		\node[draw, circle, scale=\scaleSCOT] (3) at (\x+\xInc*2,\y+\yInc*2) {$c,0$};
		\node[draw, circle, scale=\scaleSCOT] (4) at (\x,\y+\yInc) {$a,1$};
		\node[draw, circle, scale=\scaleSCOT] (5) at (\x+\xInc,\y+\yInc) {$b,1$};
		\node[draw, circle, scale=\scaleSCOT] (6) at (\x+\xInc*2,\y+\yInc) {$c,1$};
		\node[draw, circle, scale=\scaleSCOT] (7) at (\x,\y) {$a,2$};
		\node[draw, circle, scale=\scaleSCOT] (8) at (\x+\xInc,\y) {$b,2$};
		\node[draw, circle, scale=\scaleSCOT] (9) at (\x+\xInc*2,\y) {$c,2$};
		%define the clients here...
		\node[draw, rectangle, scale=\scaleBox] (S1)	at (\x+\xInc*2,\y+\yInc*2+0.7)  {$S1$};
		\node[draw, rectangle, scale=\scaleBox] (P)	at (\x+\xInc,\y-0.7) 				{$P$};
		\node[draw, rectangle, scale=\scaleBox] (S2)	at (\x+\xInc*3-0.5,\y+\yInc)	{$S2$};
		\node[draw, rectangle, scale=\scaleBox] (S3)	at (\x,\y-0.7) 					{$S3$};
		\node[draw, rectangle, scale=\scaleBox] (S4)	at (\x+\xInc*2,\y-0.7) 			{$S4$};
		% draw striaght overlay links
		\draw [gray!70, thick] (1) -- (2) (2) -- (3) (4) -- (5) (5) -- (6) (7) -- (8) (8) -- (9);
		\draw [dashed, gray!70, thick] (1) -- (4) (4) -- (7) (2) -- (5) (3) -- (6) (6) -- (9) (8) -- (5);
		\draw [thick] (P) -- (8) (S1) -- (3) (S2) -- (6) (S3) -- (7) (S4) -- (9);	
		% draw curved overlay links
		\draw [dashed, thick, gray!70]  (1) to [out=300,in=60] (7);
		\draw [line width=0.08cm, dashed, gray!70]  (2) to [out=240,in=120] (8);
		\draw [dashed, gray!70, thick]  (3) to [out=240,in=120] (9);
		%arrows of P messages
		\path [line, thick] (P) to [out=60,in=290] (8);
		\path [line, dashed, thick, red] (8) to [out=60,in=300] (5); 
		\path [line, dashed, thick] (5) to [out=60,in=300] (2); 
		\path [line, thick] (5) to [out=30,in=150] (6);
		\path [line, thick] (2) to [out=30,in=150] (3);
		\path [line, thick] (8) to [out=30,in=150] (9);
		\path [line, thick] (6) to [out=30,in=150] (S2);
		\path [line, thick] (8) to [out=150,in=30] (7);
		\path [line, thick] (3) to [out=60,in=300] (S1);
		\path [line, thick] (9) to [out=250,in=110] (S4);
		\path [line, thick] (7) to [out=250,in=110] (S3);
		\end{tikzpicture}
		\caption{\footnotesize Subscription forwarding in a cyclic overlay when load on the links is normal.}
		\label{fig:IDR2}
	\end{subfigure}
	~
	\begin{subfigure}[b]{0.23\textwidth}
		\begin{tikzpicture}
		\tikzstyle{line} = [draw, -latex']
		\centering
		\def\scaleSCOT {0.6}
		\def\scaleBox {0.7}
		\def\x {-0.5} 
		\def\xInc {1.4}
		\def\y {0} 
		\def\yInc {0.9}
		\def\xIncFirst {0.8}
		%begin the first topology diagram here ...
		\tikzstyle{every node} = [thick]
		\node[draw, circle, scale=\scaleSCOT] (1) at (\x,\y+\yInc*2) {$a,0$};
		\node[draw, circle, scale=\scaleSCOT] (2) at (\x+\xInc,\y+\yInc*2) {$b,0$};
		\node[draw, circle, scale=\scaleSCOT] (3) at (\x+\xInc*2,\y+\yInc*2) {$c,0$};
		\node[draw, circle, scale=\scaleSCOT] (4) at (\x,\y+\yInc) {$a,1$};
		\node[draw, circle, scale=\scaleSCOT] (5) at (\x+\xInc,\y+\yInc) {$b,1$};
		\node[draw, circle, scale=\scaleSCOT] (6) at (\x+\xInc*2,\y+\yInc) {$c,1$};
		\node[draw, circle, scale=\scaleSCOT] (7) at (\x,\y) {$a,2$};
		\node[draw, circle, scale=\scaleSCOT] (8) at (\x+\xInc,\y) {$b,2$};
		\node[draw, circle, scale=\scaleSCOT] (9) at (\x+\xInc*2,\y) {$c,2$};
		%define the clients here...
		\node[draw, rectangle, scale=\scaleBox] (S1)	at (\x+\xInc*2,\y+\yInc*2+0.7)  {$S1$};
		\node[draw, rectangle, scale=\scaleBox] (P)	at (\x+\xInc,\y-0.7) 				{$P$};
		\node[draw, rectangle, scale=\scaleBox] (S2)	at (\x+\xInc*3-0.5,\y+\yInc)	{$S2$};
		\node[draw, rectangle, scale=\scaleBox] (S3)	at (\x,\y-0.7) 					{$S3$};
		\node[draw, rectangle, scale=\scaleBox] (S4)	at (\x+\xInc*2,\y-0.7) 			{$S4$};
		% draw striaght overlay links
		\draw [gray!70, thick] (1) -- (2) (2) -- (3) (4) -- (5) (5) -- (6) (7) -- (8) (8) -- (9) ;
		\draw [dashed, gray!70, thick] (1) -- (4) (4) -- (7) (2) -- (5) (3) -- (6) (6) -- (9);
		\draw [thick] (P) -- (8) (S1) -- (3) (S2) -- (6) (S3) -- (7) (S4) -- (9);	
		% draw curved overlay links
		\draw [dashed, thick, gray!70]  (1) to [out=300,in=60] (7);
		\draw [line width=0.08cm, dashed, gray!70]  (2) to [out=240,in=120] (8);
		\draw [dashed, gray!70, thick]  (3) to [out=240,in=120] (9);
		\draw [line width=0.08cm, dashed, gray!70] (8) -- (5);
		%arrows of P messages
		\path [line, thick] (P) to [out=60,in=290] (8);
		\path [line, dashed, thick] (9) to [out=60,in=300] (6); 
		\path [line, dashed, thick] (9) to [out=130,in=230] (3);
		\path [line, thick, red] (8) to [out=30,in=150] (9);
		\path [line, thick] (6) to [out=30,in=150] (S2);
		\path [line, thick] (8) to [out=150,in=30] (7);
		\path [line, thick] (3) to [out=60,in=300] (S1);
		\path [line, thick] (9) to [out=250,in=110] (S4);
		\path [line, thick] (7) to [out=250,in=110] (S3);
		\end{tikzpicture}
		\caption{\footnotesize Subscription forwarding in a cyclic overlay when load on the links is normal.}
		\label{fig:IDR1}
	\end{subfigure}	
	~
	\begin{subfigure}[b]{0.23\textwidth}
		\begin{tikzpicture}
		%define line/arrow styles...
		\tikzstyle{line} = [draw, -latex']
		\centering
		\def\scaleSCOT {0.6}
		\def\scaleBox {0.7}
		\def\x {-0.5} 
		\def\xInc {1.4}
		\def\y {0} 
		\def\yInc {0.9}
		\def\xIncFirst {0.8}
		%begin the first topology diagram here ...
		\tikzstyle{every node} = [thick]
		\node[draw, circle, scale=\scaleSCOT] (1) at (\x,\y+\yInc*2) {$a,0$};
		\node[draw, circle, scale=\scaleSCOT] (2) at (\x+\xInc,\y+\yInc*2) {$b,0$};
		\node[draw, circle, scale=\scaleSCOT] (3) at (\x+\xInc*2,\y+\yInc*2) {$c,0$};
		\node[draw, circle, scale=\scaleSCOT] (4) at (\x,\y+\yInc) {$a,1$};
		\node[draw, circle, scale=\scaleSCOT] (5) at (\x+\xInc,\y+\yInc) {$b,1$};
		\node[draw, circle, scale=\scaleSCOT] (6) at (\x+\xInc*2,\y+\yInc) {$c,1$};
		\node[draw, circle, scale=\scaleSCOT] (7) at (\x,\y) {$a,2$};
		\node[draw, circle, scale=\scaleSCOT] (8) at (\x+\xInc,\y) {$b,2$};
		\node[draw, circle, scale=\scaleSCOT] (9) at (\x+\xInc*2,\y) {$c,2$};
		%define the clients here...
		\node[draw, rectangle, scale=\scaleBox] (S1)	at (\x+\xInc*2,\y+\yInc*2+0.7)  {$S1$};
		\node[draw, rectangle, scale=\scaleBox] (P)	at (\x+\xInc,\y-0.7) 				{$P$};
		\node[draw, rectangle, scale=\scaleBox] (S2)	at (\x+\xInc*3-0.5,\y+\yInc)	{$S2$};
		\node[draw, rectangle, scale=\scaleBox] (S3)	at (\x,\y-0.7) 					{$S3$};
		\node[draw, rectangle, scale=\scaleBox] (S4)	at (\x+\xInc*2,\y-0.7) 			{$S4$};
		% draw striaght overlay links
		\draw [gray!70, thick] (1) -- (2) (2) -- (3) (4) -- (5) (5) -- (6) (7) -- (8) (8) -- (9);
		\draw [dashed, gray!70, thick] (1) -- (4) (4) -- (7) (6) -- (9) (8) -- (5);
		\draw [thick] (P) -- (8) (S1) -- (3) (S2) -- (6) (S3) -- (7) (S4) -- (9);	
		% draw curved overlay links
		\draw [dashed, thick, gray!70]  (1) to [out=300,in=60] (7);
		\draw [line width=0.08cm, dashed, gray!70]  (2) to [out=240,in=120] (8);
		\draw [dashed, gray!70, thick]  (3) to [out=240,in=120] (9);
		\draw [line width=0.08cm, dashed, gray!70] (2) -- (5) (3) -- (6);
		%arrows of P messages
		\path [line, thick] (P) to [out=60,in=290] (8);
		\path [line, dashed, thick] (6) to [out=60,in=300] (3); 
		\path [line, dashed, thick,red] (8) to [out=60,in=300] (5);
		\path [line, thick,red] (5) to [out=30,in=150] (6);
		\path [line, thick] (8) to [out=30,in=150] (9);
		\path [line, thick] (6) to [out=30,in=150] (S2);
		\path [line, thick] (8) to [out=150,in=30] (7);
		\path [line, thick] (3) to [out=60,in=300] (S1);
		\path [line, thick] (9) to [out=250,in=110] (S4);
		\path [line, thick] (7) to [out=250,in=110] (S3);
		\end{tikzpicture}
		\caption{\footnotesize Subscription forwarding in a cyclic overlay when load on the links is normal.}
		\label{fig:IDR3}
	\end{subfigure}
	~
	\begin{subfigure}[b]{0.23\textwidth}
		\begin{tikzpicture}
		%define line/arrow styles...
		\tikzstyle{line} = [draw, -latex']
		\centering
		\def\scaleSCOT {0.6}
		\def\scaleBox {0.7}
		\def\x {-0.5} 
		\def\xInc {1.5}
		\def\y {0} 
		\def\yInc {0.9}
		\def\xIncFirst {0.8}
		%begin the first topology diagram here ...
		\node[draw, circle, scale=\scaleSCOT] (1c) at (\x,\y+\yInc*2) 			{$a,0$};
		\node[draw, circle, scale=\scaleSCOT] (2c) at (\x+\xInc,\y+\yInc*2) 	{$b,0$};
		\node[draw, circle, scale=\scaleSCOT] (3c) at (\x+\xInc*2,\y+\yInc*2) 	{$c,0$};
		\node[draw, circle, scale=\scaleSCOT] (4c) at (\x,\y+\yInc) 			{$a,1$};
		\node[draw, circle, scale=\scaleSCOT] (5c) at (\x+\xInc,\y+\yInc) 		{$b,1$};
		\node[draw, circle, scale=\scaleSCOT] (6c) at (\x+\xInc*2,\y+\yInc) 	{$c,1$};
		\node[draw, circle, scale=\scaleSCOT] (7c) at (\x,\y) 					{$a,2$};
		\node[draw, circle, scale=\scaleSCOT] (8c) at (\x+\xInc,\y) 			{$b,2$};
		\node[draw, circle, scale=\scaleSCOT] (9c) at (\x+\xInc*2,\y) 			{$c,2$};
		%define the clients here...
		\node[draw, rectangle, scale=\scaleBox] (S1c)	at (\x,\y+\yInc*2+0.7)  			{$S1$};
		\node[draw, rectangle, scale=\scaleBox] (Pc)	at (\x+\xInc,\y-0.7) 				{$P$};
		\node[draw, rectangle, scale=\scaleBox] (S2c)	at (\x+\xInc*3-0.5,\y+\yInc)		{$S2$};
		\node[draw, rectangle, scale=\scaleBox] (S3c)	at (\x+\xInc*2+0.4,\y-0.7) 			{$S3$};
		\node[draw, rectangle, scale=\scaleBox] (S4c)	at (\x+\xInc*2-0.3,\y-0.7) 			{$S4$};
		% draw striaght overlay links
		\draw [gray!70, thick] (1c) -- (2c) (2c) -- (3c) (4c) -- (5c) (5c) -- (6c) (7c) -- (8c);
		\draw [dashed, gray!70, thick] (1c) -- (4c) (4c) -- (7c) (2c) -- (5c) (3c) -- (6c) (6c) -- (9c);
		\draw [thick] (Pc) -- (8c) (S1c) -- (1c) (S2c) -- (6c) (S3c) -- (9c) (S4c) -- (9c);
		% draw curved overlay links
		\draw [dashed, thick, gray!70]  (1c) to [out=300,in=60] (7c);
		\draw [line width=0.08cm, dashed, gray!70]  (2c) to [out=240,in=120] (8c);
		\draw [dashed, gray!70, thick]  (3c) to [out=240,in=120] (9c);
		\draw [line width=0.08cm, dashed, gray!70] (8c) -- (5c);
		\draw [line width=0.08cm, gray!70] (8c) -- (9c);	
		%arrows of P messages
		\path [line, thick] (Pc) to [out=60,in=290] (8c);
		\path [line, dashed, thick, red] (8c) to [out=60,in=300] (5c); 
		\path [line, dashed, thick] (5c) to [out=60,in=300] (2c);
		\path [line, thick] (8c) to [out=30,in=150] (9c);
		\path [line, thick] (5c) to [out=30,in=150] (6c);
		\path [line, thick] (6c) to [out=30,in=150] (S2c);
		\path [line, thick] (1c) to [out=60,in=300] (S1c);
		\path [line, thick] (9c) to [out=210,in=90] (S4c);
		\path [line, thick] (9c) to [out=330,in=90] (S3c);
		\path [line, thick] (2c) to [out=150,in=30] (1c);
		\end{tikzpicture}
		\caption{\footnotesize Subscription forwarding in a cyclic overlay when load on the links is normal.}
		\label{fig:IDR4}
	\end{subfigure}
\end{figure*}

Notifications are routed to interested subscribers using the RPF technique, following the principle of \textit{downstream replication} \cite{carz_thesis}. A notification from a publisher may have different delivery delays depending on when the notification is received by interested subscribers. Three factors are important in determining delivery delays: (i) delay due to in-broker processing to find the next destinations, (ii) length of routing paths, and (iii) notifications payload. We introduce two algorithms for notification routing in a cluster-based SCOT: (i) \textit{Static Notification Routing (SNR)}, and (ii) \textit{Inter-cluster Dynamic Routing (IDR)}. We also compare our algorithms with state--of--the--art TID-based static and dynamic routing \cite{Li_ADAP}. 

\subsection{Static Notification Routing}
The \textit{Static Notification Routing (SNR)} algorithm uses shortest-lengths subscription trees for routing in a cluster-based SCOT (cf. Section 6). SNR algorithm deals with two scenarios: (i) when the host cluster of a publisher is the only TC of the publisher, and (ii) when the publisher has at least one STC. In (i), all interested subscribers are hosted by the brokers in the publisher's host cluster, and lengths of the routing paths satisfy the relation $max\big(d\langle(u_{1},v_{1}), (u_{2}, v_{2})\rangle \big) \leq diam(G_{af})$, where $(u_{1},v_{1})$ and $(u_{2}, v_{2})$ are any brokers in the publisher's host clusters. In this case, no inter-cluster messaging takes place as no notification is forwarded to secondary brokers and all CIBs in CIV of the advertisement of the notification sending publisher are 0 (i.e., OFF). In (ii), notification sending publisher has at least one STC, and lengths of the routing paths satisfy the relation $max\big(d\langle(x_{1},y_{1}), (x_{2}, y_{2})\rangle \big) \leq \big( diam(G_{af}) + 1 \big)$, where $(x_{1},y_{1})$ and $(x_{2}, y_{2})$ are any brokers in a SCOT overlay. As subscriptions are saved in PRTs of brokers of their host clusters, the host broker of notifications sending publisher uses CIBs of CIV in $A_{cxt}$ context to find STCs (CIB of an STC must be ON in CIV in $A_{cxt}$ context). The prime objective of allocating a bit in CIV of each advertisement is to make host broker of a publisher aware of STCs. In HC of a publisher, notifications are routed in reverse using RPF technique onto paths generated by subscription trees. Notifications are forwarded to secondary brokers of STCs using CIBs in CIV of the matching advertisement. Importantly, subscriptions of STCs are not available to host broker of the publisher that issued the matching advertisement. Recall, CIBs in CIV in $A_{cxt}$ context are used to forward matching notifications to STCs through secondary brokers of the publisher's host broker. After a broker of an STC receives a notification, RPF technique is used along subscription trees to forward notification to interested subscribers in STCs. 

\begin{figure}
	\begin{tikzpicture}
	\def\scaleSCOT {0.7}
	\def\scaleBox {0.8}
	%define line/arrow styles...
	\tikzstyle{line} = [draw, -latex']
	\tikzstyle{lineR} = [draw, latex-']
	\def\y {6}
	\def\x {0}
	\def\xInc {1.1}
	\def\yInc {1.2}		
	\node[draw, circle, scale=\scaleSCOT, pattern=north west lines, pattern color=gray!40] (1) at (\x,\y) 			{$a,0$};
	\node[draw, circle, scale=\scaleSCOT, pattern=north west lines, pattern color=gray!40] (2) at (\x + \xInc*1,\y) {$b,0$};
	\node[draw, circle, scale=\scaleSCOT, pattern=north west lines, pattern color=gray!40] (3) at (\x+\xInc*2,\y)	{$c,0$};
	\node[draw, circle, scale=\scaleSCOT, pattern=north west lines, pattern color=gray!40] (4) at (\x+\xInc*3,\y) 	{$d,0$};
	\node[draw, circle, scale=\scaleSCOT, pattern=north west lines, pattern color=gray!40] (5) at (\x+\xInc*4,\y) 	{$e,0$};
	\node[draw, circle, scale=\scaleSCOT, pattern=north west lines, pattern color=gray!40] (6) at (\x+\xInc*5+0.5,\y) 	{$f,0$};
	\node[draw, circle, scale=\scaleSCOT, fill=green!50] (11) at (\x,\y-\yInc) 			{$a,1$};
	\node[draw, circle, scale=\scaleSCOT, fill=green!50] (12) at (\x+ \xInc*1,\y-\yInc) {$b,1$};
	\node[draw, circle, scale=\scaleSCOT, fill=green!50] (13) at (\x+ \xInc*2,\y-\yInc) {$c,1$};
	\node[draw, circle, scale=\scaleSCOT, fill=green!50] (14) at (\x+ \xInc*3,\y-\yInc) {$d,1$};
	\node[draw, circle, scale=\scaleSCOT, fill=green!50] (15) at (\x+ \xInc*4,\y-\yInc) {$e,1$};
	\node[draw, circle, scale=\scaleSCOT, fill=green!50] (16) at (\x+ \xInc*5+0.5,\y-\yInc) {$f,1$};
	%draw brokers...
	\node[draw, circle, scale=\scaleSCOT] (21) at (\x,\y-\yInc*2) 		   {$a,2$};
	\node[draw, circle, scale=\scaleSCOT] (22) at (\x+ \xInc*1,\y-\yInc*2) {$b,2$};
	\node[draw, circle, scale=\scaleSCOT] (23) at (\x+ \xInc*2,\y-\yInc*2) {$c,2$};
	\node[draw, circle, scale=\scaleSCOT] (24) at (\x+ \xInc*3,\y-\yInc*2) {$d,2$};
	\node[draw, circle, scale=\scaleSCOT] (25) at (\x+ \xInc*4,\y-\yInc*2) {$e,2$};
	\node[draw, circle, scale=\scaleSCOT] (26) at (\x+ \xInc*5+0.5,\y-\yInc*2) {$f,2$};
	%draw both the clients S2 and S3
	\node[draw, rectangle, scale=\scaleBox, text=red] (S1)	at  (\x-1,\y) 					{$S1$};
	\node[draw, rectangle, scale=\scaleBox, text=red] (S2)	at (\x+ \xInc*5+1.5, \y) 		{$S2$};
	\node[draw, rectangle, scale=\scaleBox, text=blue] (S3)	at (\x+ \xInc*5+1.5, \y-\yInc) 	{$S3$};
	\node[draw, rectangle, scale=\scaleBox] (S4) at  (\x+ \xInc*5+1.5,\y-\yInc*2) 			{$S4$};
	\node[draw, rectangle, scale=\scaleBox, text=red] (P1)	at (\x-1, \y-\yInc*2) 			{$P1$};
	\node[draw, rectangle, scale=\scaleBox, text=blue] (P2)	at (\x-1, \y-\yInc) 			{$P2$};
	\node[draw, rectangle, scale=\scaleBox] (P3) at  (\x+ \xInc*5-0.2,\y-\yInc*1.5)	 		{$P3$};
	
	%draw the curved line...
	\draw [thick, gray!70] (22) to [out=25,in=155] (24) (12) to [out=25,in=155] (14) (2) to [out=25,in=155] (4);
	\draw [thick, gray!70] (1) -- (2) (2) -- (3) (4) -- (5) (5) -- (6) (11) -- (12) (12) -- (13) (14) -- (15) (15) -- (16) (21) -- (22) (22) -- (23) (24) -- (25) (25) -- (26);
	%draw the dashed i-COL lines...					
	\draw  [dashed, thick, gray!70] (1) -- (11) (11) -- (21) (2) -- (12) (12) -- (22) (3) -- (13) (13) -- (23) (4) -- (14) (14) -- (24) 
	(5) -- (15) (15) -- (25) (6) -- (16) (16) -- (26);
	%draw the curved dotted links...		
	\draw [dashed, thick, gray!70] (1) to [out=240,in=120] (21) (2) to [out=240,in=120] (22) (3) to [out=240,in=120] (23) (4) to [out=240,in=120] 
	(24) (5) to [out=240,in=120] (25) (6) to [out=240,in=120] (26);
	%draw msgs from P3.
	\draw [line, thick, dotted] (P3) to [out=170,in=310] (15);
	\draw [line, thick, dotted] (15) to [out=30,in=160] (16); 
	\draw [line, thick, dotted] (16) to [out=30,in=160] (S3);
	\draw [line, thick, dotted] (15) to [out=60,in=300] (5); 
	\draw [line, thick, dotted] (15) to [out=290,in=70] (25);
	\draw [line, thick, dotted] (25) to [out=30,in=150] (26);
	\draw [line, thick, dotted] (26) to [out=40,in=140] (S4);
	\draw [line, thick, dotted] (5) to [out=30,in=160] (6); 
	\draw [line, thick, dotted] (6) to [out=30,in=160] (S2);
	\draw [line, thick, dotted] (5) to [out=150,in=30] (4); 
	\draw [line, thick, dotted] (4) to [out=140,in=35] (2); 
	\draw [line, thick, dotted] (2) to [out=150,in=25] (1); 
	\draw [line, thick, dotted] (1) to [out=150,in=25] (S1); 	 	
	%draw msgs from P2.
	\draw [line, thick, blue, dashed] (P2) to [out=15,in=165] (11);
	\draw [line, thick, blue, dashed] (11) to [out=60,in=300] (1); 
	\draw [line, thick, blue, dashed] (11) to [out=300,in=60] (21);
	\draw [line, thick, blue, dashed] (21) to [out=330,in=210] (22); 
	\draw [line, thick, blue, dashed] (22) to [out=335,in=205] (24);
	\draw [line, thick, blue, dashed] (24) to [out=335,in=205] (25);
	\draw [line, thick, blue, dashed] (25) to [out=335,in=205] (26);
	\draw [line, thick, blue, dashed] (26) to [out=335,in=205] (S4);
	\draw [line, thick, blue, dashed] (1) to [out=210,in=335] (S1);	
	%draw msgs from P1.
	\draw [line, thick] (26) to [out=15,in=165] (S4);
	\draw [line, thick] (21) to [out=15,in=165] (22);
	\draw [line, thick] (P1) to [out=15,in=165] (21);
	\draw [line, thick] (25) to [out=15,in=165] (26);
	\draw [line, thick] (24) to [out=15,in=165] (25);
	\draw [line, thick] (22) to [out=30,in=150] (24);
	
	\draw  (S1) -- (1) (S2) -- (6); 
	\draw  (S3) -- (16); 
	\draw (S4) -- (26) (P1) -- (21) (P2) -- (11) (P3) -- (15); 
	\end{tikzpicture}
	\captionof{figure}{\small \textit{The two-step subscription forwarding process in the SCOT shown in Figure 3. The solid arrows indicate part of the subscription tree generated in the first step and the dashed arrows indicate part of the subscription tree generated in the second step. }}
	\label{fig:theNFP}
\end{figure}
The SNR algorithm is further explained in Fig. \ref{fig:theNFP}, which shows publisher \textit{P1} is hosted by B(a,2) in $C_{2}$, while \textit{P2} and \textit{P3} are hosted by B(a,1), B(e,1) in $C_{1}$, respectively. The sets of subscribers interested in notifications from P1, P2, and P3 are \textit{\{S4\}}, \textit{\{S1, S4\}} and \textit{\{S1, S2, S3, S4\}}. The set of TCs of P1, P2 and P3 are $\{C_{2}\}, \{C_{0}, C_{2}\}$ and $\{C_{0},C_{1},C_{2}\}$. CIVs (in $A_{cxt}$ context) of advertisements of P1, P2, and P3 are 100, 101, and 111, respectively.

Inter-cluster routing of notifications is achieved without storing subscriptions in PRTs of a subscriber's secondary brokers. A notification from P1 propagates onto the path indicated by the solid black arrow in $C_{2}$ to reach S4. As P1 has no STCs, no CIB in CIV of the advertisement of P1 has value 1 and notification from P1 propagates only onto aLinks. 

In case of P2, no notification propagate onto aLinks as the HC of P2 is not its target cluster. B(a,1), the host broker of P2, creates two copies of each notification from P2 and forwards each copy onto links $\langle(a,1), (a,0) \rangle$ and $\langle(a,1), (a,2) \rangle$. P2 has only STCs because no broker in its HC is interested in receiving notifications from P2. Notifications routing paths from P2 to S1 and S3 are indicated by the dashed blue arrow messages. Brokers in all three clusters are interested in notifications from P3. B(e, 1), the host broker of P3, creates three copies of a notification from P3 and uses CIV of advertisement of P3 to propagate notifications onto links $\langle(e,1), (e,2) \rangle$, and $\langle(e,1), (e,0) \rangle$. Another copy of the notification is forwarded onto aLink $\langle(e,1), (f,1) \rangle$ due to subscription tree of S3. S1 and S2 are interested in notifications from P3, only one copy of the notification is forwarded to B(e,0) to avoid sending duplicates.

\subsection{Inter-cluster Dynamic Routing}
\textit{Dynamic Routing} refers to the capability of a PS system to alter the routing path in response to overloaded or failed links and/or brokers. Dynamic routing requires multiple paths from a publisher to interested subscribers. Multiple techniques have been proposed to handle dynamic routing in address-based networks where routing paths are calculated from a global view of a network topology graph that is saved on every network router \cite{routing_book}. However, these techniques are not applicable in the PS domain, because brokers are aware of only their direct neighbours and dynamic routing decisions have to be made without having a global view of an overlay and without making changes in content-based routing paths due to performance reasons. This paper focuses on \textit{Inter-cluster Dynamic Routing (IDR)} of notifications when one or more iLinks are overloaded and notifications start queuing up. This can happen in two cases: (i) when a broker is not able to process high volume of outgoing notifications and become overwhelm, and (ii) when bandwidth is limited. 

There is always one advertisement-tree per advertisement, which is used by a matching subscription to generate a unique routing path for notification forwarding. This makes IDR a challenge in APS system. We use the \textit{structuredness} of a cluster-based SCOT to perform IDR when one or more iLinks are overloaded. IDR can reduce delivery delays when a large number of notifications start accumulating in the output queues of a broker. We use Eq. 10 to find whether an output queue is congested, and the IDR should find alternative iLinks. We have used the same equation for our previous work for subscription-based dynamic routing \cite{OctopiS}. 
\begin{equation}
(Q_{\ell}) . (1+Q_{in}, 1+Q_{out})_{t_{w}}  > \tau 
\end{equation}
(need to twist words in this paragraph...) $Q_{in}$ is the number of notifications that enter into the output queue, and $Q_{out}$ is the number of notifications that leave the output queue in time window $t_{w}$. The term $(1+Q_{in}, 1+Q_{out})_{t_{w}}$ is the ratio of $(1+Q_{in})$ to $(1+Q_{out})$, and is known as the \textit{Congestion Element (CE)}. $CE > 1$ indicates that the congestion is increased in the last $t_{w}$ interval as more messages entered into the output queue than sent out onto the overlay link to next broker. $CE < 1$ shows that congestion is decreased in the last $t_{w}$ interval, as more messages are sent out onto the overlay link than entered into the output queue. $Q_{\ell}$ is the length of the output queue (i.e., the number of notifications waiting in the output queue to be forwarded to next destinations). An output queue is congestion--free when $CE$ is 1 and $Q_{\ell}$ is 0. OctopiA saves the values of $Q_{in}$ and $Q_{out}$ in a \textit{Link Status Table (LST)} on each broker, and the values are updated after each $t_{w}$ interval. IDR in an APS system becomes active when Eg. 10 is valid for some positive real value of $\tau$ in $t_{w}$ interval. In the SNR algorithm, a broker adds exclusive copies of a notification in the output queues of the target links. For example, if a publisher generates $\gamma$ number of notifications in $t_{w}$ interval, and there are $\alpha$ target aLinks and $\beta$ target iLinks at a broker in notification routing path, then the broker adds an $(\alpha + \beta).\gamma$ number of notifications in the output queues of the target links in $t_{w}$ interval. A High Rate Publisher (HRP) with a very high value of $\gamma$ can overwhelm the host broker and the brokers in notification routing paths. IDR algorithm can alleviate overwhelmed brokers by adding fewer copies of a notification in the congested output queues of the target iLinks. In particular, IDR adds \textit{no} copy of a notification in the congested output queues when at least one output queue of a target aLink or iLink is uncongested. Furthermore, when the output queues of all target links are congested, IDR adds only one copy of the notification to the least congested output queue of a target iLink. The relation between $|iLinks_{ol}|$ and $|V_{cf}-1|$, where $|iLinks_{ol}|$ is the number of overloaded target iLinks at a broker, provides information about unoverloaded target iLinks that can be used to forward a notification. The mechanism of IDR is further explained by using three different cases. Fig. 7 shows a SCOT whose $G_{af}$ is a \textit{path graph} of three nodes and $G_{cf}$ is a triangle. The set of subscribers \textit{\{S1, S2, S3, S4\}} is interested in notifications from a publisher P.  

(i) When at least one target iLink is unoverloaded, then the relation $|iLinks_{ol}| < |V_{cf}-1|$ is valid. The output queues of $|iLinks_{ol}|$ target iLinks are congested and at least one target iLink is unoverloaded. The notification routing broker sets the CIBs of overloaded target iLinks to 1. Important to mention that the CIBs of a target iLink and an STC, whose broker is connected to notification routing broker through that target iLink, are same. The context of CIV is notification routing is always $P_{cxt}$. No notification is added to the congested output queue of target iLinks. Instead, CIV is added to the header of the notification which is added into the output queue of an unoverloaded target iLink. The notification which carries CIV is called \textit{the CIV-Notification (or CIV-N)}. If there are multiple unoverloaded target iLinks, CIV-N is added into output queue of the target iLink which has the least value of $Q_{\ell}$. The number of notifications added to the output queues of the target links in $t_{w}$ interval is $(\alpha + \theta )\gamma$, where $\theta$ is the number of unoverloaded target iLinks and $\theta < \beta$. Using this technique, IDR algorithm keeps the length of the congested output queues of a broker of the target iLinks short. The load of forwarding the notification to STCs is shifted to the next primary broker using the heuristic that unoverloaded target iLinks are available down the routing path. 

CIV-N notifications are distinguished by the red arrow messages in Fig. 7. Fig. 7(a) indicates that the iLinks, $l \langle(b,2), (b,0) \rangle$ is overloaded (thick dashed link), and IDR algorithm forwards a notification from P to S3 and S4 in the host cluster. Since $l \langle(b,2), (b,1) \rangle$ target iLink is unoverloaded, CIV-N notification (with CIV 001) is passed into that link. As $l \langle(b,1), (b,0) \rangle$ is unoverloaded, B(b,1) forwards a copy of the notification to B(b,0) and another to B(c,1). To forward a notification from P in this case, the IDR and SNR algorithm generate 6 IMs, while routing paths generated by IDR contain one extra broker. However, IDR eliminates the overloaded iLink $l \langle(b,2),(b,0)\rangle$ from the routing paths, which is not possible with the SNR algorithm.

(ii) (how many notifications are added) The condition $|iLinks_{ol}| = |V_{cf}-1|$ at any broker in a routing oath indicates that all the target iLinks are overloaded. We assume that at least one target aLink is unoverloaded, which forwards CBV-N to next destination. If more than one unoverloaded target aLinks are available, CBV-N is forwarded onto the target aLink with the least value of $Q_{\ell}$. The number of notifications added to the output queues of the target links in $t_{w}$ interval is $\alpha .\gamma$. Fig. 7(b) indicates that the CBV-N, with CBV 011, is forwarded to $l \langle(b,2),(c,2)\rangle$) using the subscription tree of S4 because all the target iLinks on B(b,2) are overloaded and (presumably), the value of $Q_{\ell}$ for the output queue of aLink $l \langle(b,2),(c,2)\rangle$ is less than $l \langle(b,2),(a,2)\rangle$. Since the target iLinks on B(c,2) are unoverloaded, a copy of the notification is forwarded onto iLinks of STCs $C_{1}$ and $C_{0}$. To forward one notification from P to interested subscribers, the IDR algorithm generates 4 IMs, while the SNR algorithm generates 5 IMs. Further, dynamic routing path generated by IDR algorithm eliminates overloaded target iLinks thus increases system's performance and throughput. The algorithm requires processing of the notification by only 4 brokers, which is much less than 6 brokers required by SNR.

(iii) (how many notifications are added) This case discuses a scenario when target iLinks on multiple clusters may be overloaded. Fig. 7(c) shows that the CIV-N message is forwarded onto the target link $l \langle(b,2),(b,1)\rangle$ with CIV 001 as the target link $l \langle(b,2),(b,0)\rangle$ is overloaded. Since $l \langle(b,1),(b,0)\rangle$ is also overloaded, the CIV-N message is passed on to B(c,1) using subscription tree of S2. At B(c,1), the only target link that is available to forward the notification to any broker of $C_{0}$ is $l \langle(c,1),(c,0)\rangle$. However, the last target iLink is also overloaded and no other unoverloaded iLink could be searched (as no unoverloaded target aLink is available at B(c,1)). Therefore, the overloaded target iLink $l \langle(c,1),(c,0)\rangle$) is used to forward notification to S1. To forward one notification from P to interested subscribers, the IDR algorithm generates 5 IMs, while the SNR algorithm generates 6 IMs. Further, dynamic routing path generated by IDR algorithm eliminates overloaded target iLinks at the publisher's host cluster but added another overloaded target iLink to send notification to S1. The IDR algorithm requires processing of the notification by 6 brokers, which is less than 7 brokers required by SNR.

(iv) When the condition $\big(|iLinks_{ol}| = |V_{cf}-1|\big)$ is valid at any broker in a notification routing path and no unoverloaded target aLink is available, the CBV-N is forwarded onto the least overloaded target iLink. The number of notifications added to the output queues of the target links in $t_{w}$ interval is $(\alpha + 1)\gamma$. Fig. 7(d) shows that CBV-N is forwarded onto the least overloaded link $l \langle(b,2),(b,1)\rangle$ with CBV is 001. The number of IMs generated by SNR and DNR algorithms is 5; however, the dynamic route used to send the notification to S1 has one additional broker. The principle of downstream replication is followed when the notification is forwarded onto the link $l\langle(b,2),(c,2)\rangle$. 

IDR is a best-effort algorithm for inter-cluster dynamic routing and depends on the subscription trees laid-on by interested subscribers, which are hosted by the routing cluster. The algorithm does not guarantee finding an unoverloaded iLink, even if one exists. In Fig. 7(c), IDR does not use the unoverloaded link $l\langle(b,2),(a,2)\rangle$ for dynamic routing because the link is not a target aLink. IDR also does not support \textit{intra-cluster dynamic routing}. An algorithm which covers both these cases will be presented in an extended version of this paper. 
\section{Evaluation}
		% trim={<left> <lower> <right> <upper>}
		\begin{figure*}
			\centering
			\begin{subfigure}[b]{0.23\textwidth}
				\includegraphics[trim=3cm 5cm 2cm 8cm, width=\textwidth]{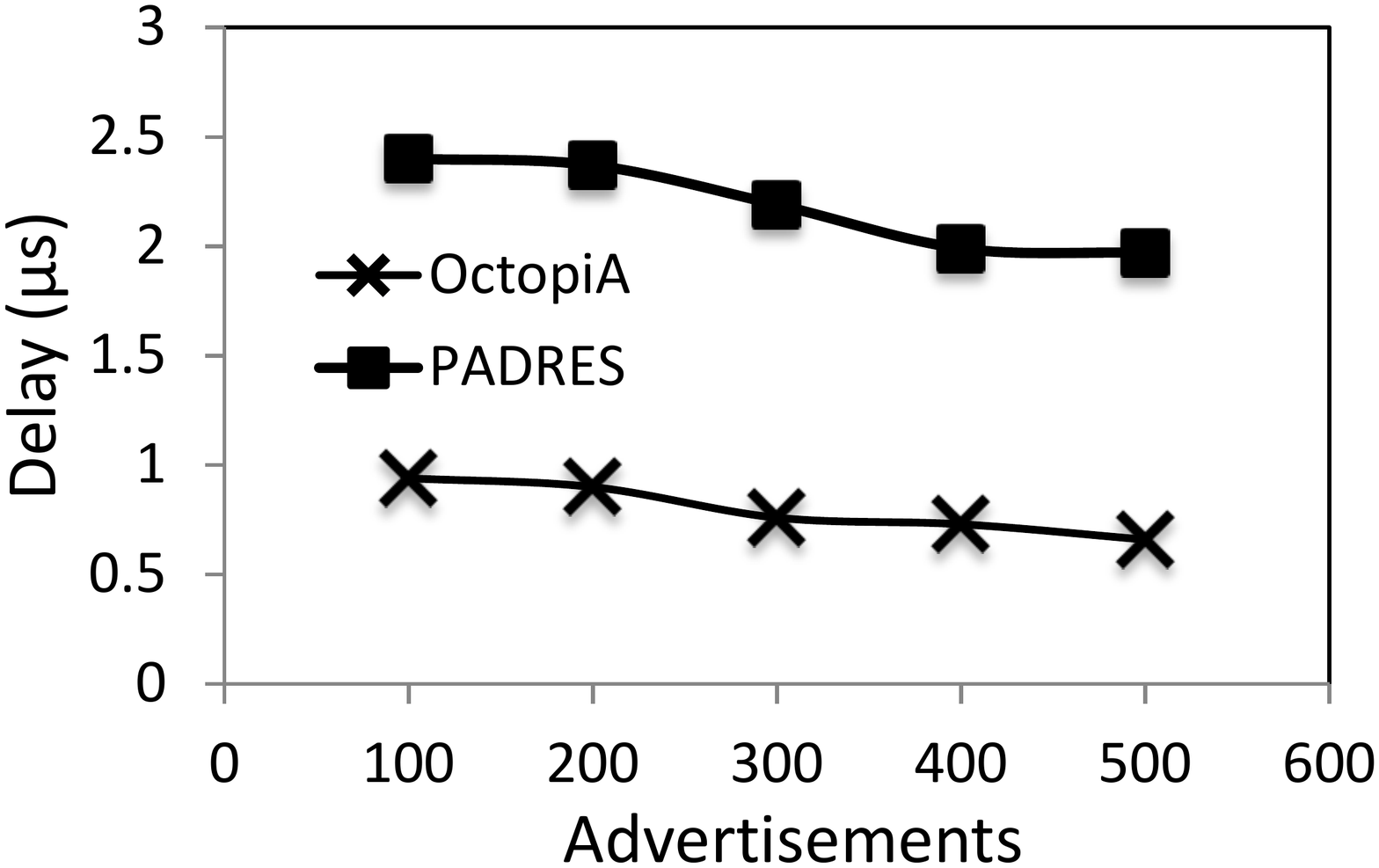}
				\caption{\footnotesize Advertisement Delay.}
				\label{fig:S1}
			\end{subfigure}
			~	%(or a blank line to force the subfigure onto a new line)
			\begin{subfigure}[b]{0.24\textwidth}
				\includegraphics[trim=3cm 5cm 2cm 8cm, width=\textwidth]{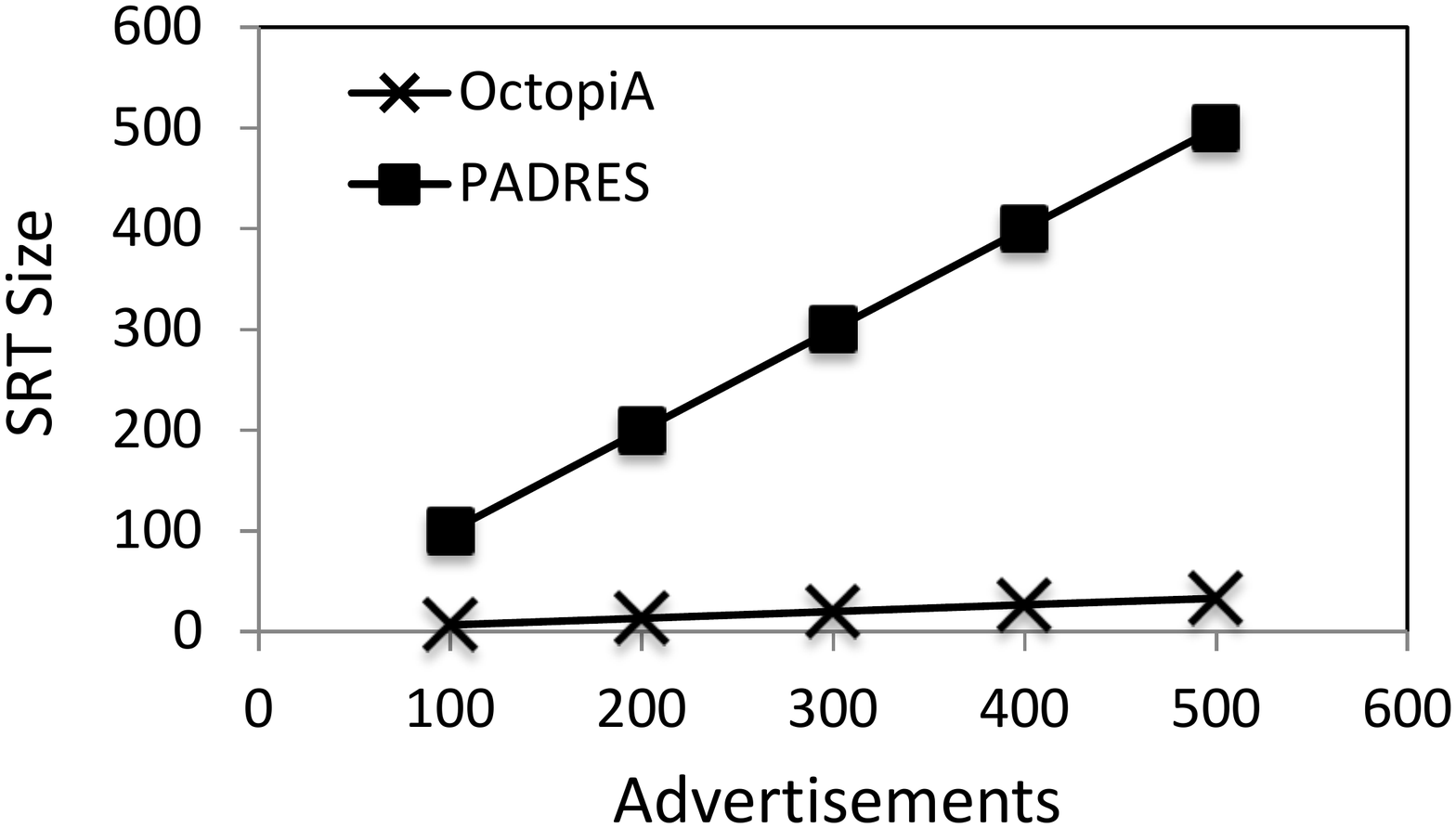}
				\caption{\footnotesize Size of SRTs.}
				\label{fig:S2}
			\end{subfigure}
			~ % start the next figure...
			\begin{subfigure}[b]{0.23\textwidth}
				\includegraphics[trim=3cm 5cm 2cm 8cm, width=\textwidth]{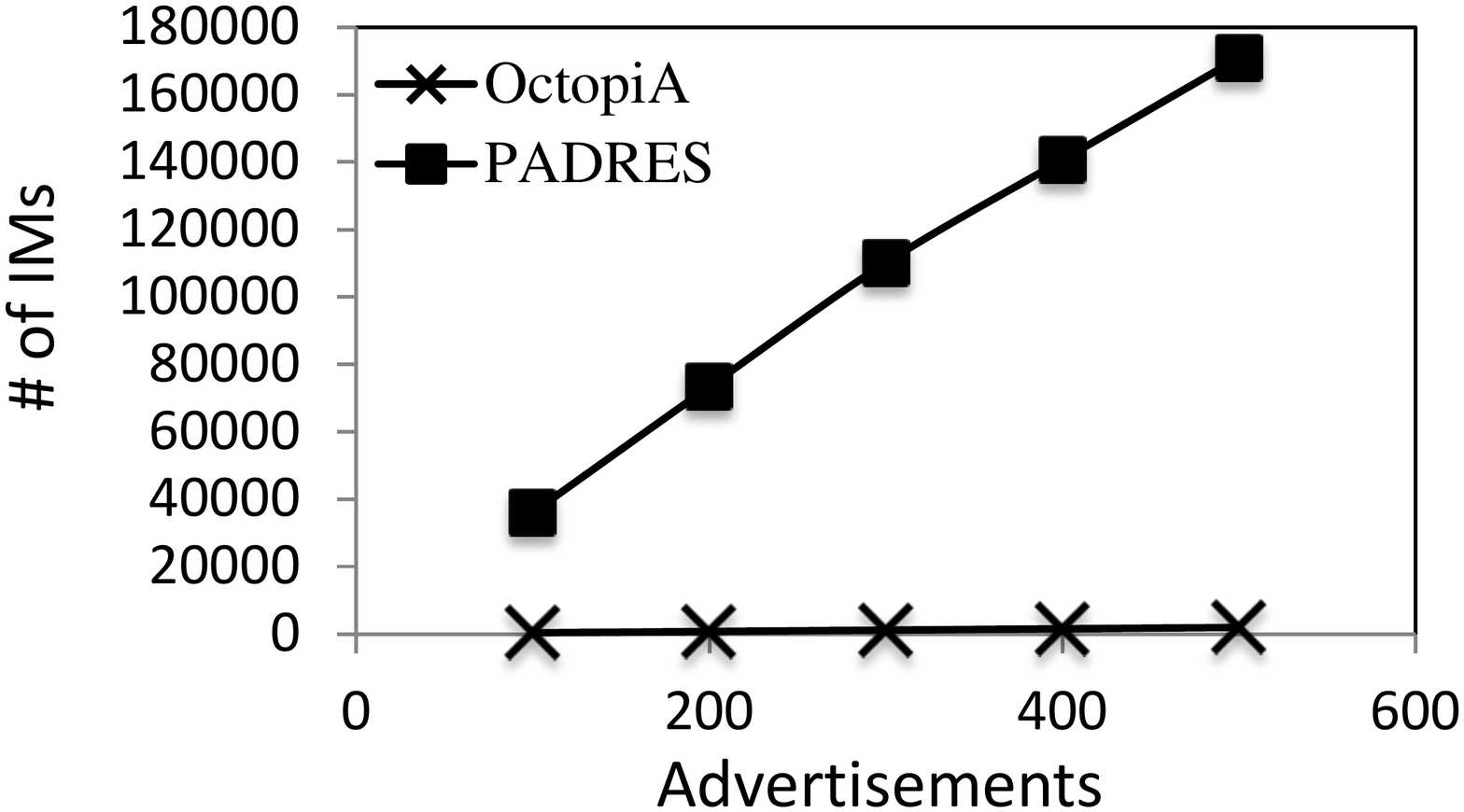}
				\caption{\footnotesize \# of IMs in AFP.}
				\label{fig:P1}
			\end{subfigure}
			\caption{\small \textit{Advertisements forwarding/broadcasting in unclustered (for PADRES) and cluster-based SCOT (for OctopiA).}}\label{fig:publication}
		\end{figure*}
		We implemented the SNR, and IDR routing algorithms in OctopiA, which is developed on top of PADRES, an open source PS tool \cite{PADRESBookChapte}. PADRES messages are serialized Java objects, which consists of two parts: the message header and the message body. OctopiA uses the language model and matching logic of PADRES, but features like patterns and sequences of events are not supported. PADRES offers content-based routing in acyclic and cyclic (general) overlays, however, OctopiA supports only structured cyclic overlays (i.e., SCOT). An unstructured cyclic overlays can have any number of brokers, and each broker can have any number of links. These properties may provide flexibility in the deployment of a PS system. However, the routing algorithms require a notification to carry TIDs of the matching advertisements. We used PADRES with an unclustered SCOT to evaluate and compare TID-based static and dynamic routing algorithms described in \cite{Li_ADAP}.
		
		\textbf{Setup ---}
		Fig. 10 shows the factors of the SCOT topology we used for evaluation of OctopiA. The $G_{af}$ factor of $\mathbb{S}$ is an acyclic topology of 14 brokers and the $G_{cf}$ factor of $\mathbb{S}$ has 5 brokers. The $G_{af}$ factor has 4 inner brokers (\textit{F, G, H} and \textit{I}) and 10 edge brokers. Since each of these brokers has  $|V_{cf}|-1$ number of secondary neighbours, there are 20 inner brokers and 50 edge brokers for a total of 70 brokers in the evaluation topology, which was deployed on a cluster of 6 computing nodes. Each computing node consisted of 2 Intel(R) Xeon(TM) E5-2620 CPU with total 12 cores of 2.1 GHz each, and 64 GB RAM. One Mellanox SX1012 switch with 10 Gbps links was used to connect the computing nodes. The topology was deployed in such a way that the primary and the secondary neighbours of each broker were always on different computing nodes. Each broker was loaded in a separate instance of JVM with 1 GB initial memory.
		\begin{tikzpicture}[thick,scale=1.2]
		\scriptsize
		\def\scaleSCOT {1}
		%define line/arrow styles...
		\def\y {6}
		\def\x {0}
		\def\xInc {0.8}
		\def\yInc {0.7}
		\node[draw, circle, minimum size=0.6cm] (1) at (\x + \xInc*0,\y) {$A$};
		\node[draw, circle, minimum size=0.6cm] (2) at (\x + \xInc*1,\y) {$B$};
		\node[draw, circle, minimum size=0.6cm] (3) at (\x + \xInc*2,\y) {$C$};
		\node[draw, circle, minimum size=0.6cm] (4) at (\x + \xInc*3,\y) {$D$};
		\node[draw, circle, minimum size=0.6cm] (5) at (\x - \xInc,\y-\yInc) {$E$};
		\node[draw, circle, minimum size=0.6cm] (6) at (\x + \xInc*0,\y-\yInc) {$F$};
		\node[draw, circle, minimum size=0.6cm] (7) at (\x + \xInc*1,\y-\yInc) {$G$};
		\node[draw, circle, minimum size=0.6cm] (8) at (\x + \xInc*2,\y-\yInc) {$H$};
		\node[draw, circle, minimum size=0.6cm] (9) at (\x + \xInc*3,\y-\yInc) {$I$};
		\node[draw, circle, minimum size=0.6cm] (10) at (\x + \xInc*4,\y-\yInc) {$J$};
		\node[draw, circle, minimum size=0.6cm] (11) at (\x + \xInc*0,\y-\yInc*2) {$K$};
		\node[draw, circle, minimum size=0.6cm] (12) at (\x + \xInc*1,\y-\yInc*2) {$L$};
		\node[draw, circle, minimum size=0.6cm] (13) at (\x + \xInc*2,\y-\yInc*2) {$M$};
		\node[draw, circle, minimum size=0.6cm] (14) at (\x + \xInc*3,\y-\yInc*2) {$N$};
		
		\draw [thick] (1) -- (6) (2) -- (7) (3) -- (8) (4) -- (9) (5) -- (6) (6) -- (7) (7) -- (8) (8) -- (9) (9) -- (10);
		\draw [thick] (11) -- (6) (12) -- (7) (13) -- (8) (14) -- (9);
		%draw the operator box.
		\node[draw, rectangle, minimum size=0.3cm] (op)    at (3.8,\y-\yInc) {};
		%draw the connectivity triangle...
		\node[draw, circle, minimum size=0.6cm] (c0) at (5.1,\y-0.1)	{$0$};
		\node[draw, circle, minimum size=0.6cm] (c1) at (4.4, \y-\yInc) {$1$};
		\node[draw, circle, minimum size=0.6cm] (c2) at (5.8,\y-\yInc) {$4$};
		\node[draw, circle, minimum size=0.6cm] (c3) at (4.7,\y-\yInc*2) {$2$};
		\node[draw, circle, minimum size=0.6cm] (c4) at (5.5,\y-\yInc*2) {$3$};
		%draw edges...
		\draw [dashed] (c0) -- (c1) (c0) -- (c2) (c0) -- (c3) (c0) -- (c4) (c1) -- (c3) (c1) -- (c2) (c1) -- (c4) (c3) -- (c4) (c3) -- (c2) (c4) -- (c2);
		\end{tikzpicture}
		
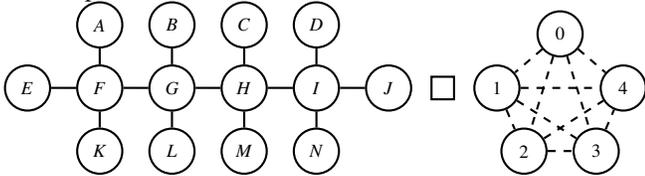
\captionof{figure}{\small \textit{Left operand of $\Box$ operator is the $G_{af}$ factor while the right operand is the $G_{cf}$ factor of $\mathbb{S}$. The $G_{cf}$ generates five replicas of $G_{af}$.}}
		\label{fig:evalTop}
		Stock datasets are commonly used to generate workloads for evaluations of content-based PS systems \cite{agg15}. We used a dataset of 500 stock symbols from Yahoo Finance!, and each stock notification had 10 distinct attributes. This high dimension data require high computation for filtering information during in-broker processing. Advertisements and matching subscriptions were generated synthetically with 2\% selectivity. We randomly distributed publishers and subscribers to brokers and each subscriber registered only one subscription.\\
		\textbf{Metrics ---}
		Through experiments with real world data, we evaluated OctopiA using primitive metrics, such as advertisement, subscription, notification, and matching delays. We also measure size of SRTs and PRTs.\\
		\textit{Advertisement Delay}: In traditional APS systems, the advertisement delay is the maximum time elapsed by an advertisement to reach every broker in an overlay network. In OctopiA, an advertisement is expected to take less time because it is forwarded to only brokers in a region and generates advertisement tree of length 1 (cf. Section ?). Since advertisement trees generated in a cluster-based SCOT are shorter than unclustered SCOT, it is important to measure the difference between the average advertisement delays in the both cases.\\ 
		\textit{Subscription Delay}: In traditional APS systems, the subscription (forwarding) delay is the maximum time elapsed as a subscription is forwarded from the subscriber, that issued the subscription, to brokers hosting publishers that issued matching advertisements. In OctpiA, subscriptions are broadcast to clusters without considering matching advertisements. The comparison between average subscription delays in two cases is important to understand subscription forwarding.\\
		\textit{Notification Delay}: The notification delay measures end-to-end latency from the time a notification is generated by a publisher to the time it is received by a subscriber. Notification delay in traditional APS systems is measured due to routing paths generated by advertisement trees and matching subscriptions. However, in OctopiA, notification is forwarded using subscription tress generated at each cluster level. The inter-cluster routing of notifications is achieved by using CIVs of the matching advertisements and without broadcasting subscriptions to all clusters of SCOT topology. Knowing the difference in latencies in these two cases is a worthwhile area of inquiry.\\
		\textit{Matching Delay}: The matching delay is the time taken by each broker in a routing path to find subscriptions that match with the notification contents. A notification in TID-based routing is matched at the host brokers of a publisher and its interested subscribers.\\
		\textit{Inter-broker Messages (IMs)}: In traditional APS systems, the number of IMs depend on the lengths of routing paths generated by advertisement trees and matching subscriptions. However, in OctopiA, cluster level subscription trees are used to forward matching notifications where inter-cluster routing is achieved by using CIVs saved with the advertisements. Investigating the number of IMs generated in PADRES, when unclustered SCOT is used, and OctopiA is worthwhile.\\
		\textit{Size of SRT}: SRT of an overlay broker saves routing information of advertisements in the form of \textit{(advertisement, lastHop)}
		entries \cite{PADRESBookChapte}. A large of routing entries cause large matching and forwarding delays in routing information. Techniques like covering and merging are used to reduce the number of entries in PRTs. However, these techniques are mainly developed for acyclic overlays and not applicable to structured cyclic overlays like SCOT (use-book-ref). These techniques do not accommodate bit-vector mechanism like CIV and are not considered in this paper. Advertisements in OctopiA are forwarded to only brokers in a SCOT region, which reduces size of SRTs dramatically. Size of SRT metric is used for this purpose.\\
		\textit{Size of PRT}: PRT of an overlay broker saves subscriptions that match with an advertisement. In traditional APS systems, routing information is saved in the form of \textit{(subscription, lastHop)} entries \cite{PADRESBookChapte}. However, subscriptions are broadcast in SCOT clusters without considering matching advertisements. Techniques like covering and merging are not applied due to same reasons described for advertisements in the last paragraphs. 
		\\ 
		% trim={<left> <lower> <right> <upper>}
		\begin{figure*}
			\centering
			\begin{subfigure}[b]{0.23\textwidth}
				\includegraphics[trim=2cm 4cm 2cm 6.7cm, width=\textwidth]{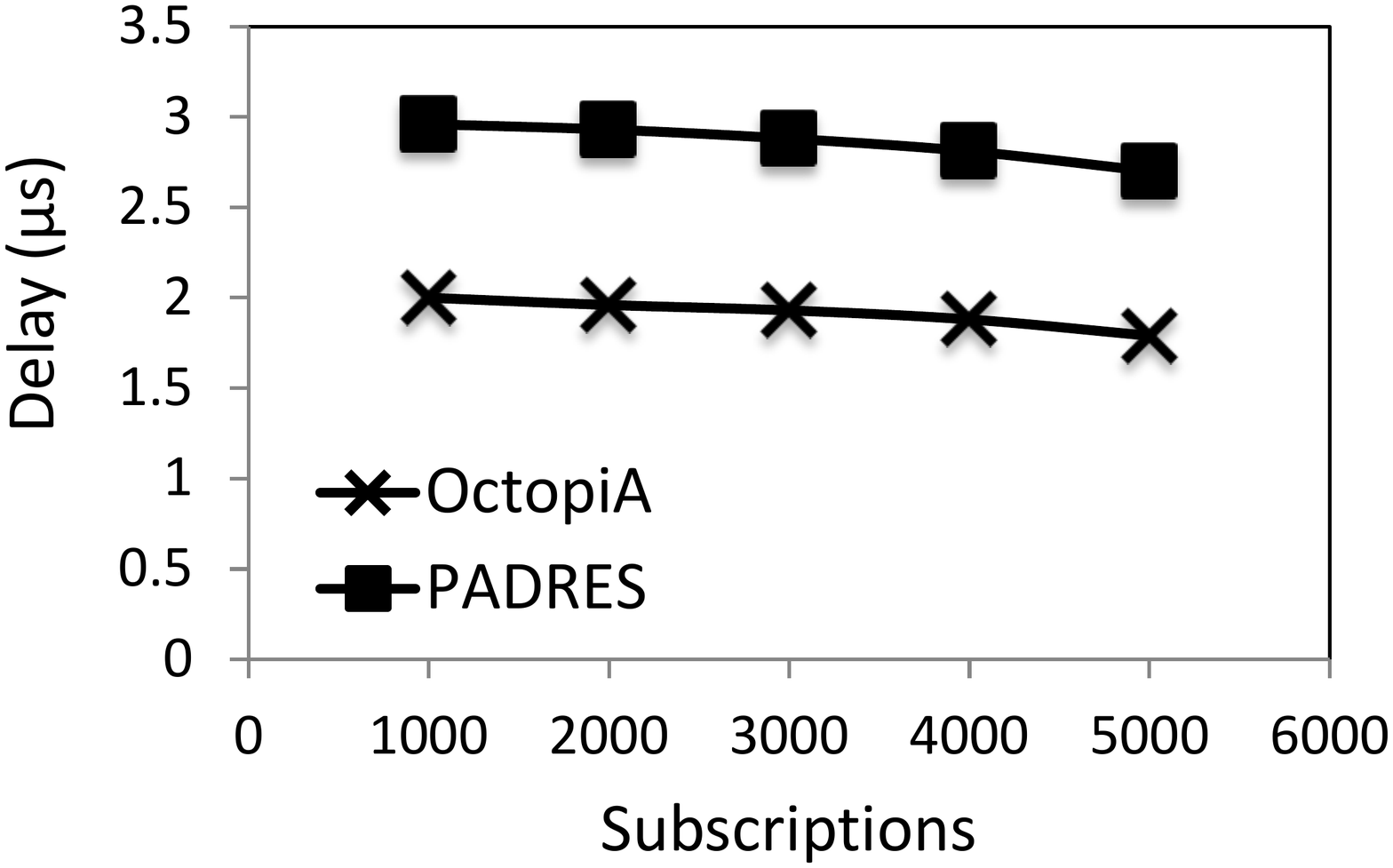}
				\caption{\footnotesize Matching Delay}
				\label{fig:S5}
			\end{subfigure}
			~
			\begin{subfigure}[b]{0.23\textwidth}
				\includegraphics[trim=3cm 4cm 2cm 6.7cm, width=\textwidth]{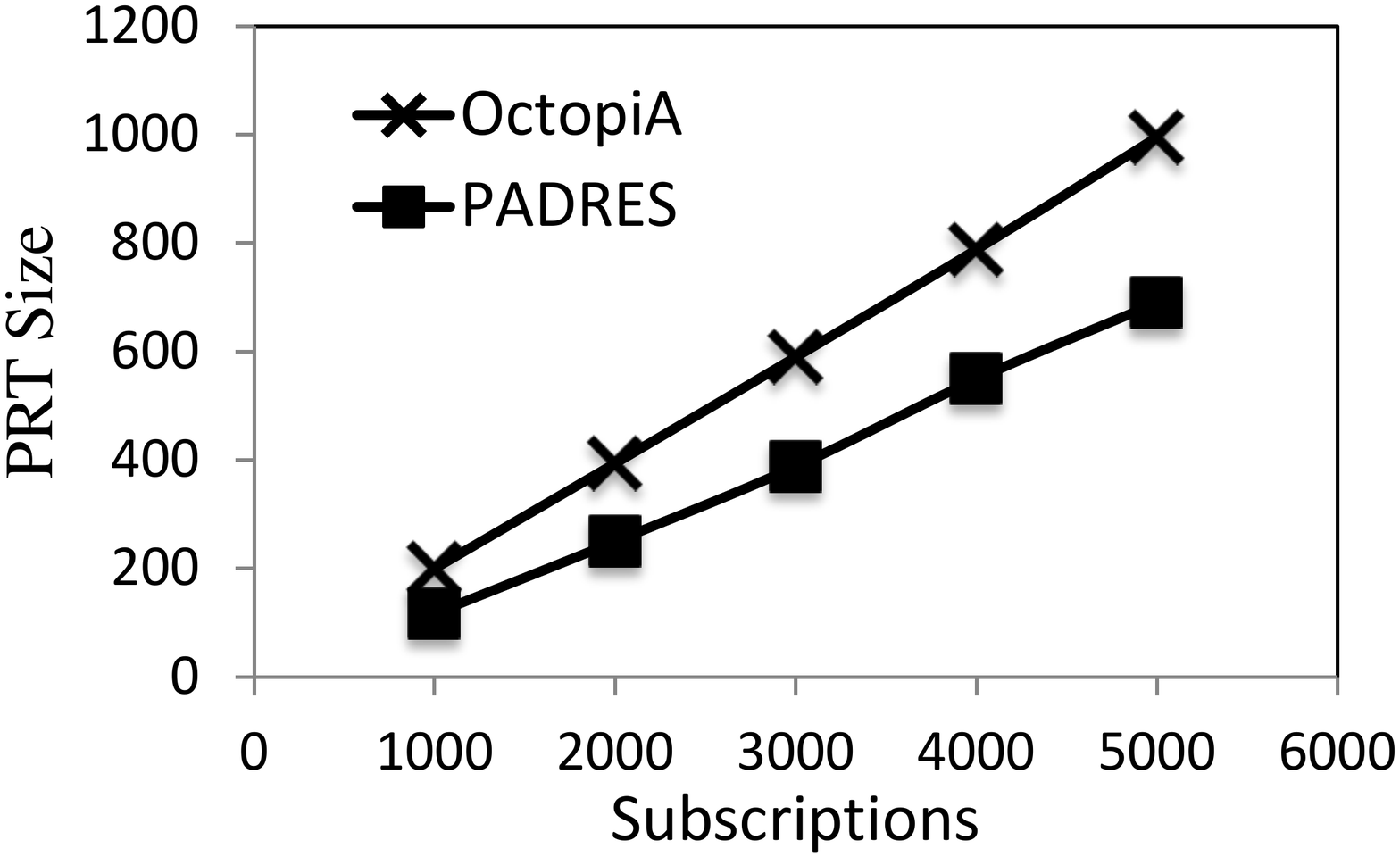}
				\caption{\footnotesize Publication Delay}
				\label{fig:P6}
			\end{subfigure}
			~ %add desired spacing between images, e. g. ~, \quad, \qquad, \hfill etc.
			%(or a blank line to force the subfigure onto a new line)
			\begin{subfigure}[b]{0.23\textwidth}
				\includegraphics[trim=3cm 4cm 2cm 6.7cm, width=0.98\textwidth]{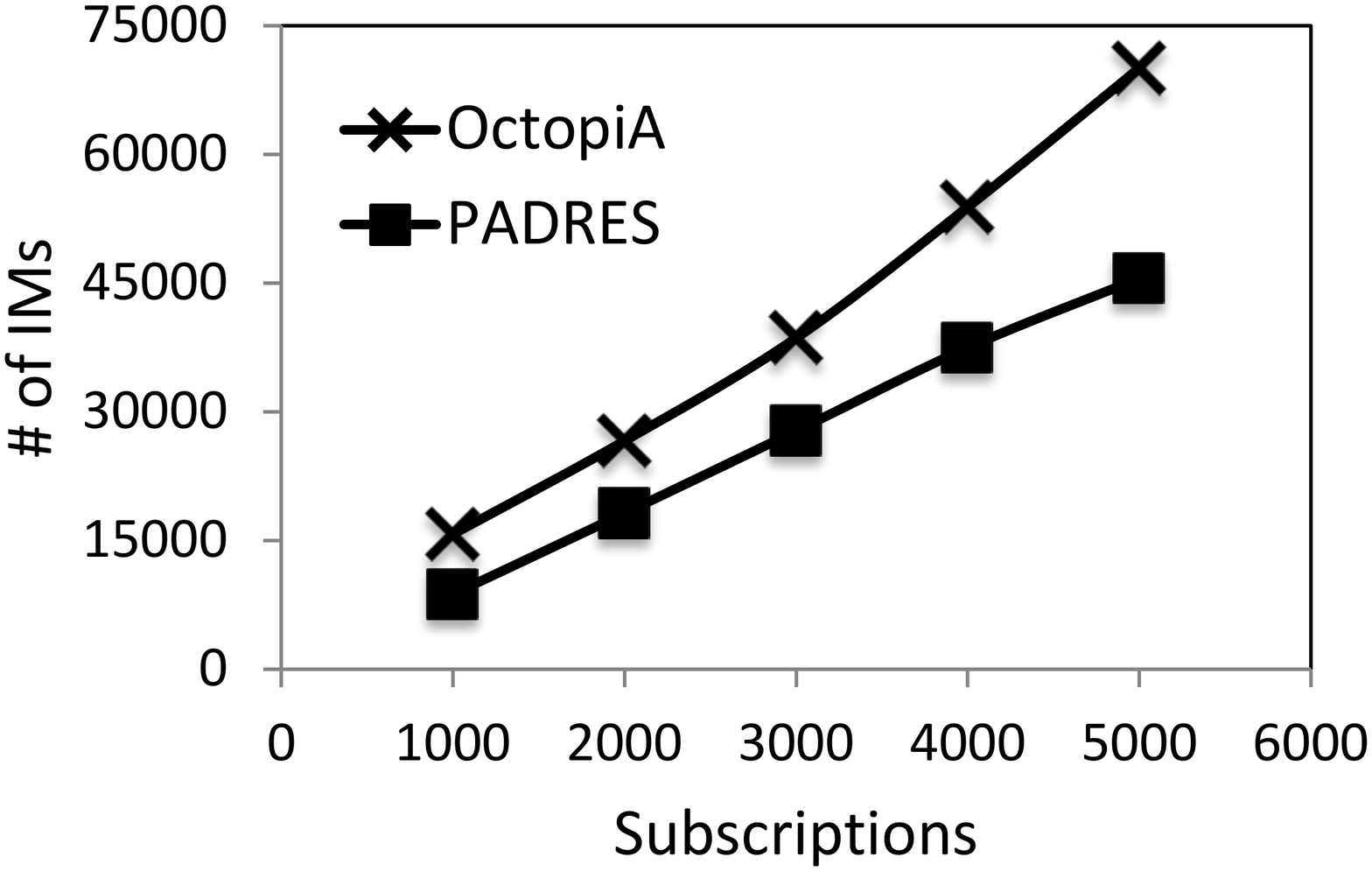}
				\caption{\footnotesize Inter-Broker Messages}
				\label{fig:P8}
			\end{subfigure}
			\caption{\small \textit{(a) Matching dealy in Subscriber Scalability. (b),(c),(d) Publisher Scalability in an unclustered and cluster-based SCOT.}}
			\label{fig:publicationNumber}
		\end{figure*}
		\begin{figure*}
			\centering
			\begin{subfigure}[b]{0.23\textwidth}
				\includegraphics[trim=2cm 4cm 2cm 5cm, width=\textwidth]{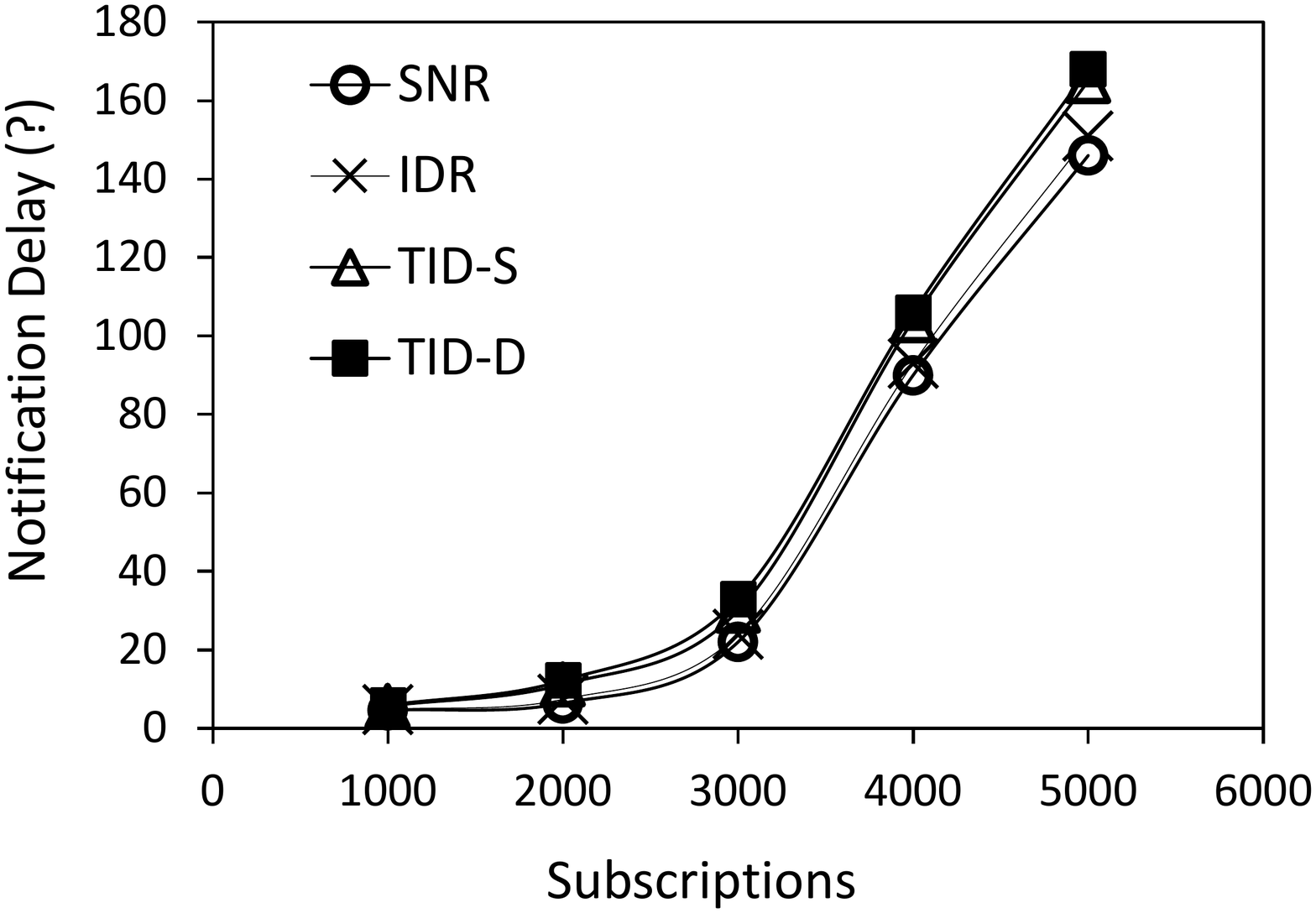}
				\caption{\footnotesize Matching Delay}
				\label{fig:S55}
			\end{subfigure}
			~
			\begin{subfigure}[b]{0.23\textwidth}
				\includegraphics[trim=3cm 4cm 2cm 5cm, width=\textwidth]{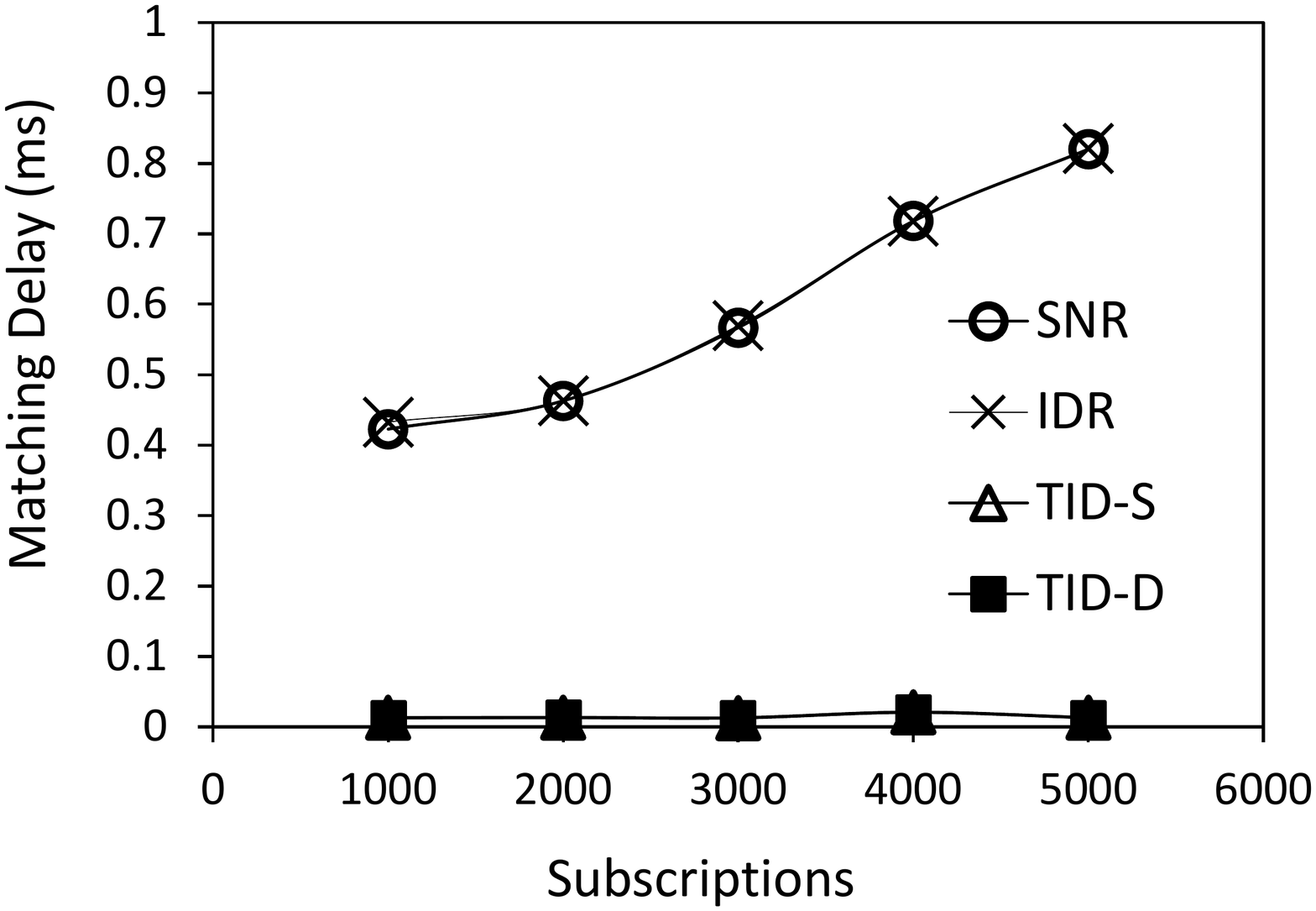}
				\caption{\footnotesize Publication Delay}
				\label{fig:P4}
			\end{subfigure}
			~ %add desired spacing between images, e. g. ~, \quad, \qquad, \hfill etc.
			%(or a blank line to force the subfigure onto a new line)
			\begin{subfigure}[b]{0.23\textwidth}
				\includegraphics[trim=3cm 4cm 2cm 5cm, width=0.98\textwidth]{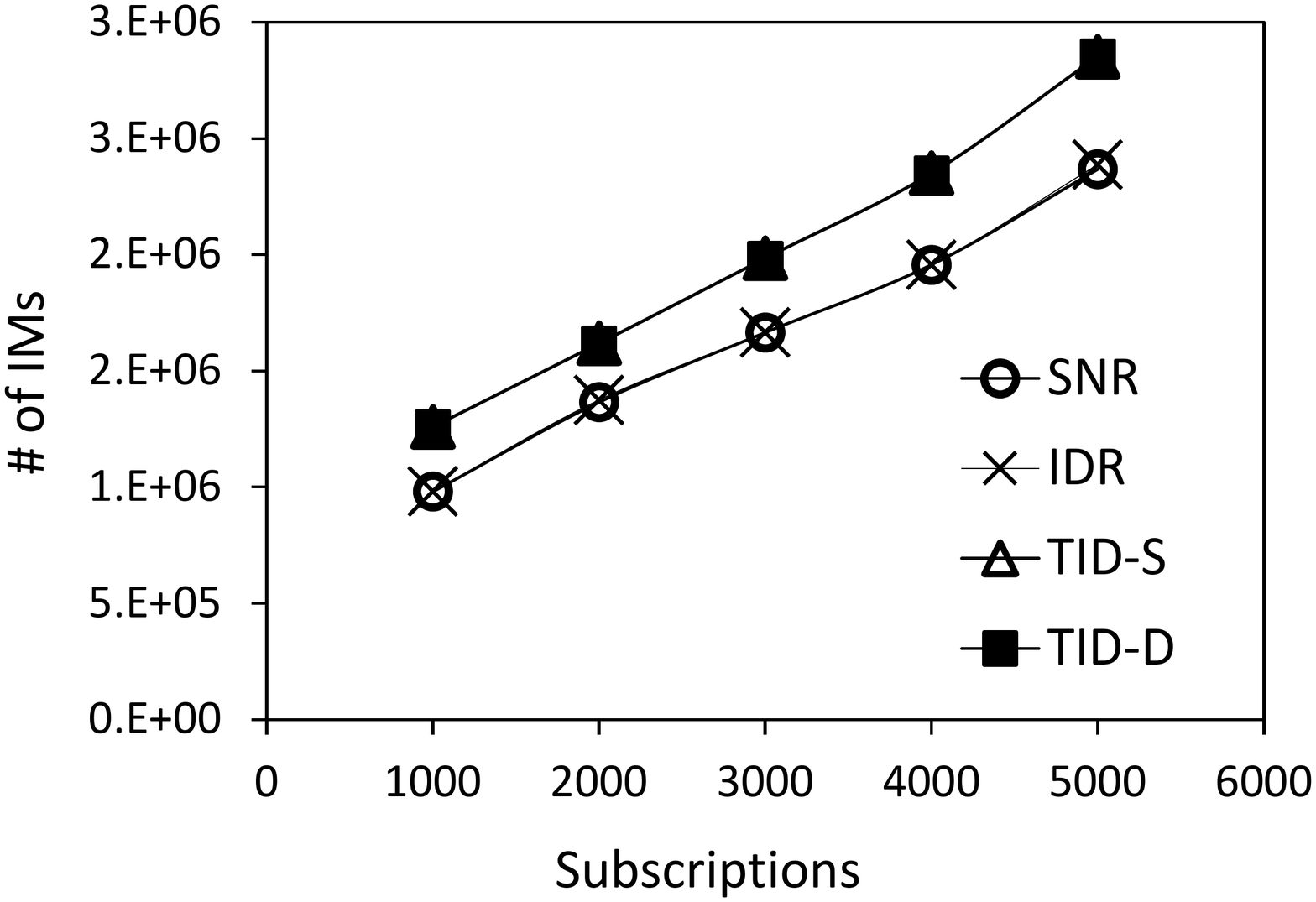}
				\caption{\footnotesize Inter-Broker Messages}
				\label{fig:P5}
			\end{subfigure}
			\caption{\small \textit{(a) Matching dealy in Subscriber Scalability. (b),(c),(d) Publisher Scalability in an unclustered and cluster-based SCOT.}}
			\label{fig:pubNumber}
		\end{figure*}
		\textbf{Results ---}
		The results presented in this section cover three important aspects of evaluation: (i) Publisher Scalability (take details to respective subheadings ), which studies the impact of increase in the number of advertisements on delivery delay, size of SRTs, and number of generated IMs; (ii) Subscriber Scalability, which studies effects like subscription delay, size of PRTs, and the number of IMs generated in subscriptions forwarding or broadcast process; (iii) End to End Notification delivery, which studies delivery and matching delays, and number of IMs generated; (iii) Burst Scenario, in which an HRP starts sending notifications at a high rate and causes congestion in the output queues. Publishers and subscribers were randomly distributed among brokers of the topology.\\
		\begin{figure*}
			\centering
			\begin{subfigure}[b]{0.23\textwidth}
				\includegraphics[trim=3cm 3.5cm 2cm 5cm, width=\textwidth]{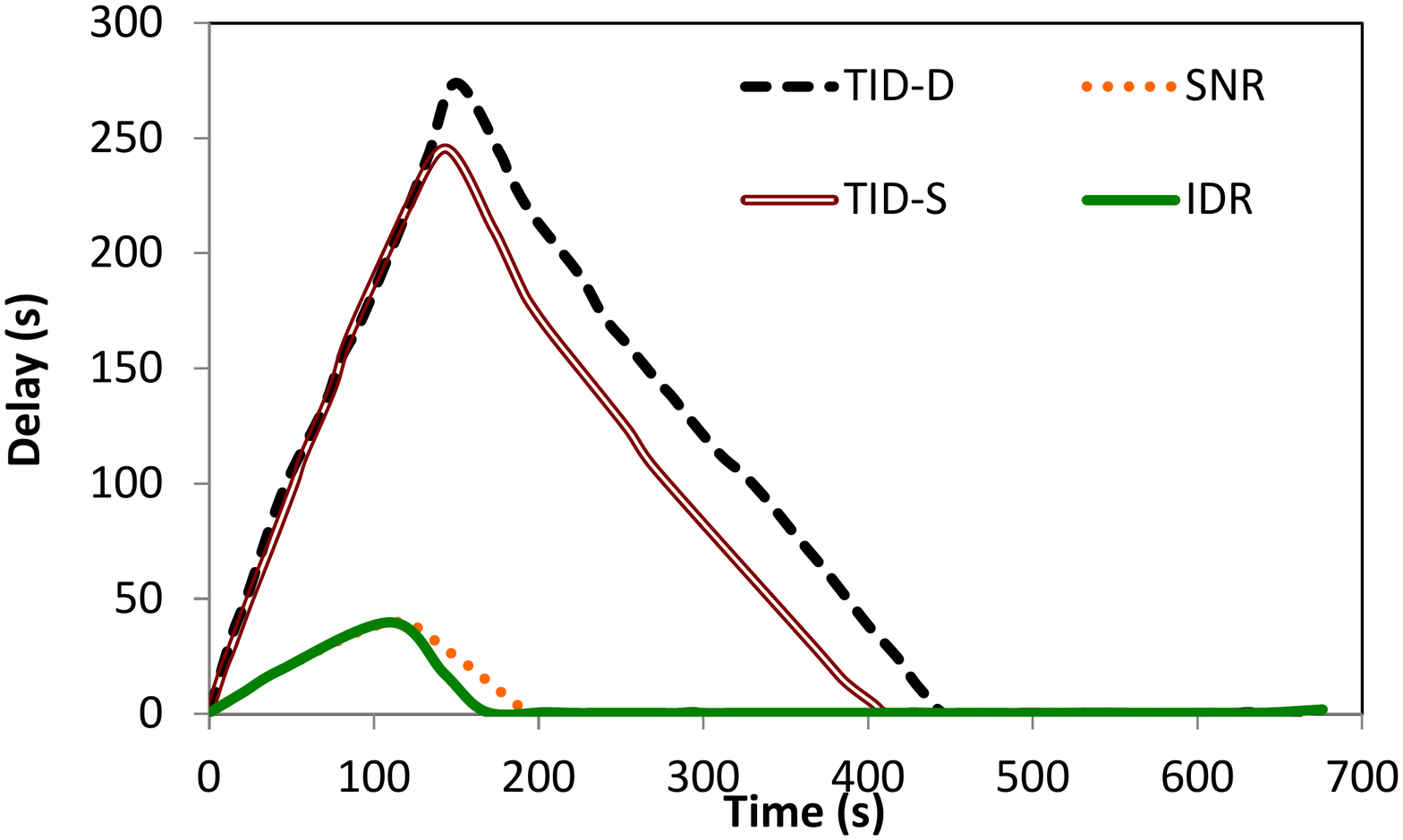}
				\caption{\footnotesize 100K messages per minute}
				\label{fig:P14}
			\end{subfigure}
			~
			\begin{subfigure}[b]{0.23\textwidth}
				\includegraphics[trim=3cm 3cm 2cm 5cm, width=\textwidth]{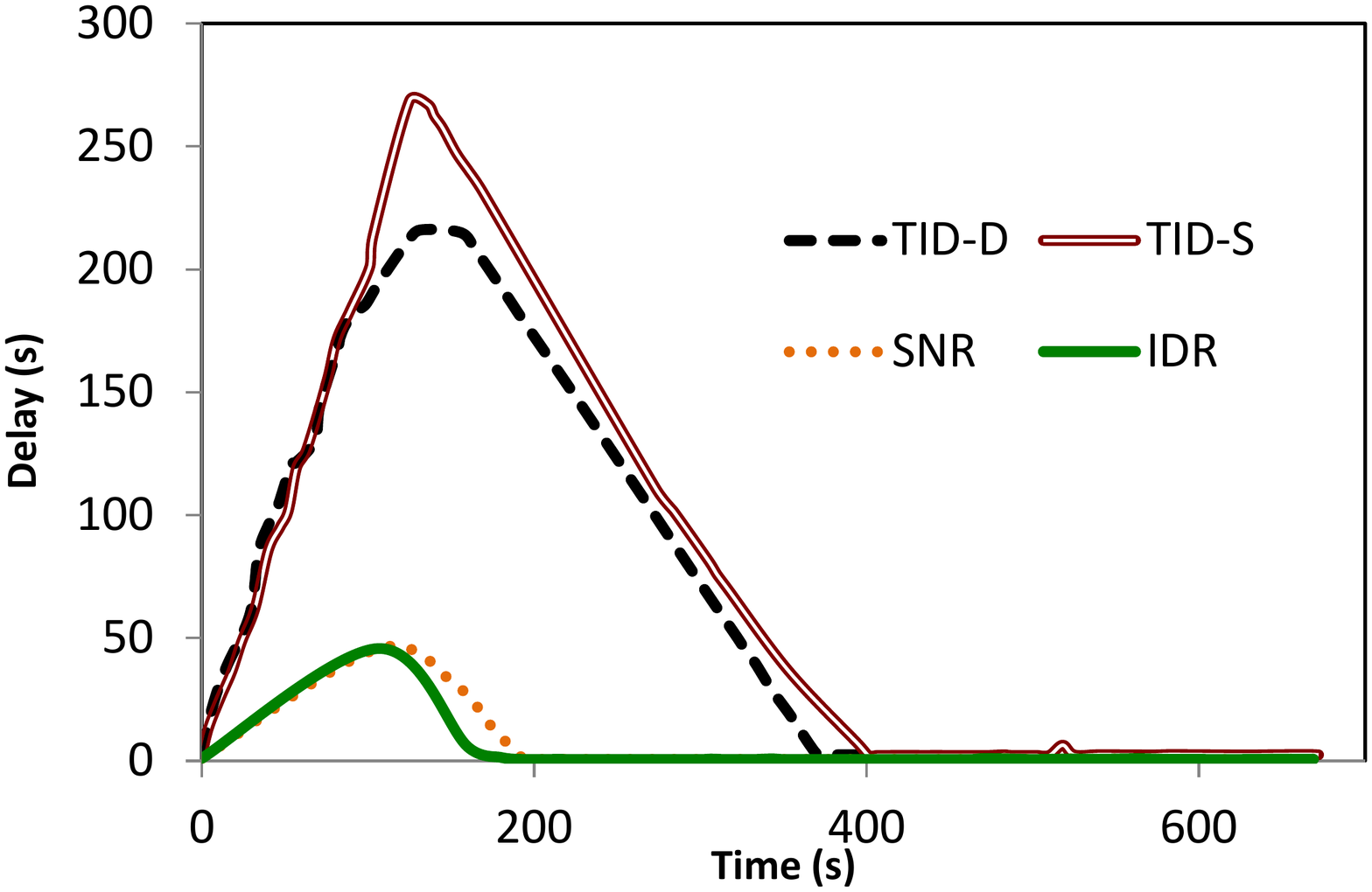}
				\caption{\footnotesize 80K messages per minute}
				\label{fig:P15}
			\end{subfigure}
			~
			\begin{subfigure}[b]{0.23\textwidth}
				\includegraphics[trim=3cm 3.5cm 2cm 5cm, width=\textwidth]{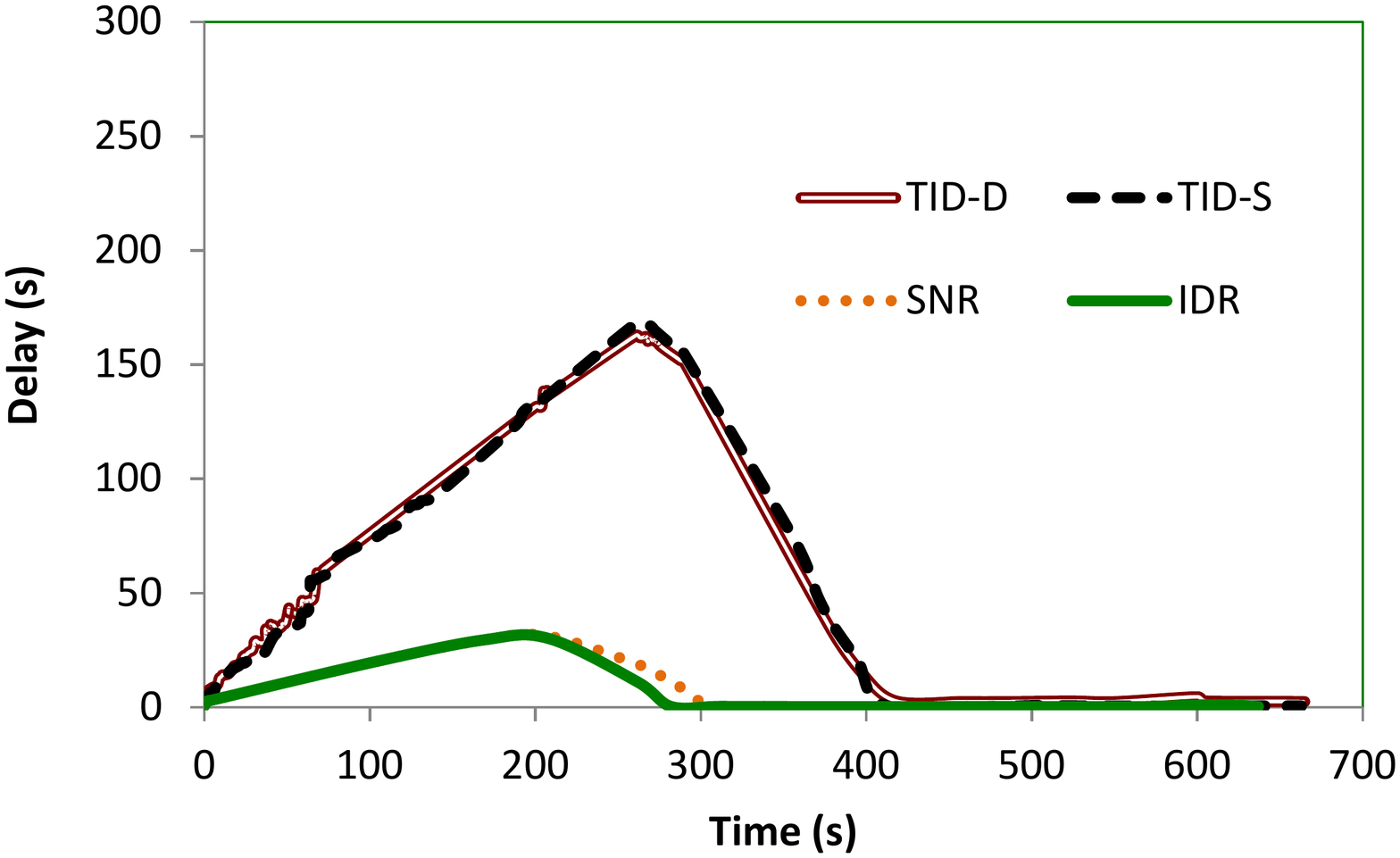}
				\caption{\footnotesize 60K messages per minute}
				\label{fig:P7}
			\end{subfigure}
			~
			~
			\begin{subfigure}[b]{0.23\textwidth}
				\includegraphics[trim=3cm 3.5cm 2cm 5cm, width=\textwidth]{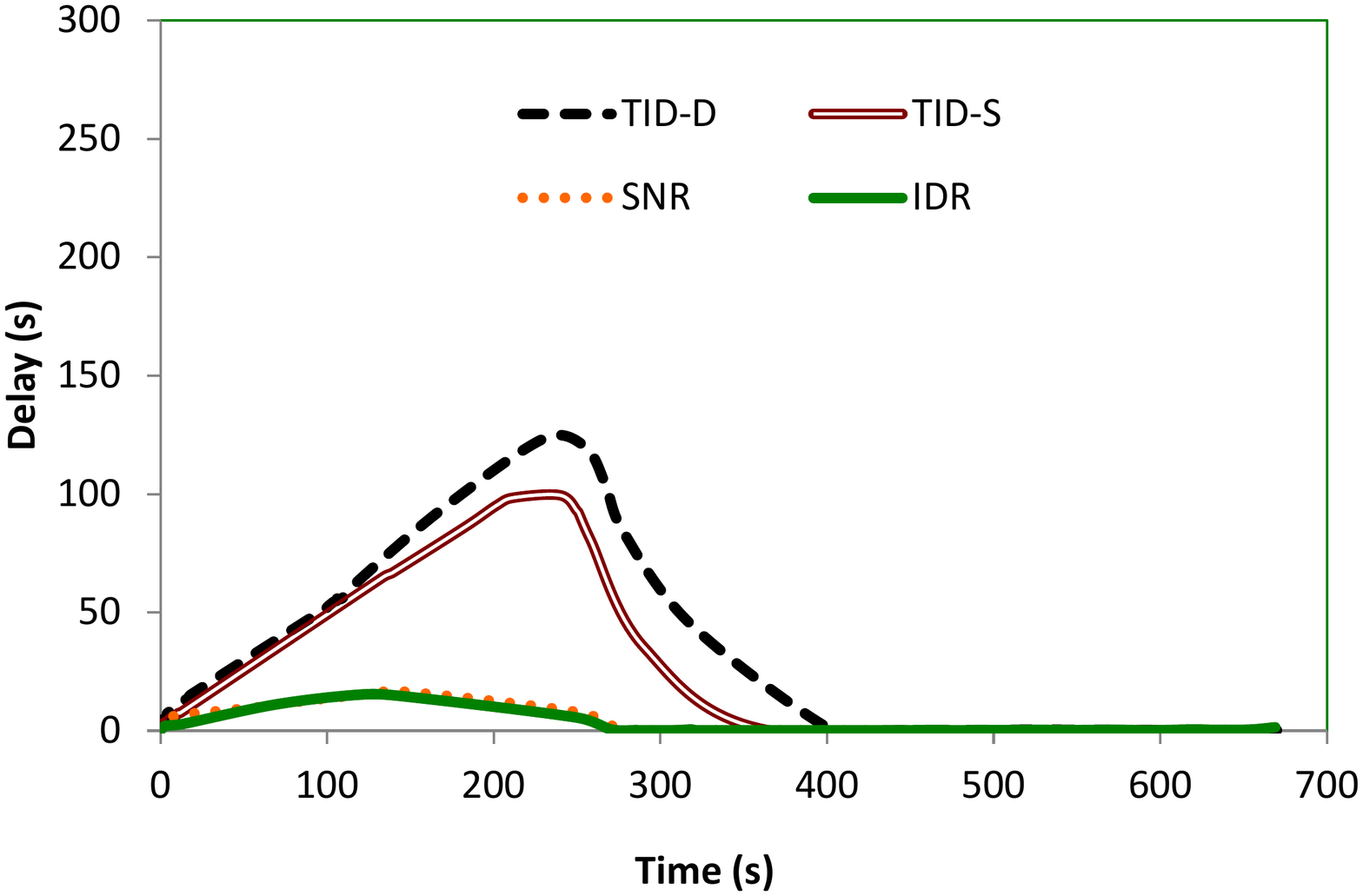}
				\caption{\footnotesize 40K messages per minute}
				\label{fig:P77}
			\end{subfigure}
			\caption{\small \textit{Dynamic routing analysis.}}
			\label{fig:burst}
		\end{figure*}
		\textit{Publisher Scalability: }
		We increased the number of advertisement from 100 to 500 in advertisement forwarding/broadcast process. Fig. 9(a) shows the average advertisement delay in unclustered SCOT (legend PADRES) is higher than the cluster-based SCOT (legend OctopiA). The figure indicates that the average advertisement delay in unclustered and cluster-based $\mathbb{S}$ is nearly constant. However, the average advertisement delay in OctopiaA is 63.5\% less than PADRES with unclustered SCOT. This difference is mainly due to the larger lengths of advertisement trees generated by advertisement broadcast algorithm used in PADRES, which can generate advertisement trees of lengths 1. The algorithm generated redundant messages to detect duplicates, while OctopiA generates advertisements trees of length 1. Fig. 9(b) shows the average sizes of SRTs when advertisements are broadcast to brokers in overlay and to brokers in one region. The average size of SRT in OctopiA is 93\% less than PADRES. The reasons are the lengths of advertisement trees generated due to advertisement broadcast algorithms. Fig. 9(c) shows that the number of IMs generated in advertisement broadcast in unclustered $\mathbb{S}$ is less than cluster-based $\mathbb{S}$. More specifically, OctopiA generates 98.9 times less number of IMs when compared with PADRES. There are two reasons for this large difference: (i) higher average lengths of advertisement trees in unclustered $\mathbb{S}$, and (ii) extra IMs generated to discard duplicates (cf. Section II).\\
		\textit{Subscriber Scalability: }
		We increased the number of subscribers from 500 to 5000, while keeping the number of publishers 100. Each subscriber issued one subscription, while each publisher issued one advertisement. Fig. 11(b) shows that the average size of PRTs in PADRES is smaller than OctopiA. More specifically, the number of entries in PRTs in OctopiA are 33\% more than PADRES. PADRES forwards subscriptions only to overlay links generated by matching advertisements. However, OctopiA broadcasts a subscription in a SCOT cluster to form subscription subgroups. The subscription broadcast approach of OctopiA generated more IMs than PADRES. Fig. 11(c) shows PADRES generated 32.7\% less IMs than OctopiA. Although the average size of PRTs and IMs are less in PADRES, Fig. 11(a) shows that subscription delay is OctopiA is 33\% less than PADRES. The key reason of this difference is matching delay in PADRES as subscriptions are forwarded after finding matching advertisements. In OctopiA, subscriptions are broadcast at cluster level and size of SRTs is much smaller than PADRES.\\
		\textit{Dynamic Notification Routing:}
		The dynamic notification routing analysis tells how quickly a CPS system converge to a normal state after an HRP finishes sending notifications. To study this behaviour, we set the value of $\tau$ to 10 and $t_{w}$ to 50 milliseconds. We used 2000 subscribers and 100 publishers in this experiment. Each subscriber had 2\% selectivity, while each publisher issued 2000 notifications at the rate of 60 per minute rate. 2\% of the subscribers subscribed to receive notifications from an HRP. We executed three simulations in which the HRP sent 100K notifications at rates of 100K, 80K, and 60K notifications per minute. The HRP and its interested subscribers did not connect to the same broker and all clusters of $\mathbb{S}$ were TCs of the HRP. This arrangement makes sure that each cluster receives notifications from the HRP. Furthermore, the HRP and its interested subscribers were hosted by different brokers exerting more load on iCOLs and aCOLs. The burst of notifications started after all the subscriptions were registered in $\mathbb{S}$. The burst continued from 60 to 100 seconds depending on the rate of the HRP. Each point in the graphs in Fig. 11 is a maximum delivery delay of 1000 notifications received in a sequence. This approach helps in analysing the stability of a SCPS system without graphing too many points. Each simulation was run until all notifications (6 million in total) were received by subscribers. Fig. 11 shows that DNR stabilized Octopi before the SNR algorithm, while the BID-based routing algorithm was not able to stabilize the system for the same workload. On average, in the first 18 minutes and 30 seconds, the maximum delay of the notification (out of 1000) was the same in the three routing algorithms. No tendency toward stability was observed. This indicates that, due to the high rates of notifications from the HRP, the state of the system (Octopi) did not converge to normal until the condition $CE < 1$ was maintained for some time (on average, 16 minutes and 40 seconds for the three simulations). DNR started stabilizing Octopi before other two algorithms. The average value of $Q_{\ell}$ of target links at the host broker of HRP when DNR was used was 61\% less than SNR and 73\% less than BID-based routing algorithms. There were 5 TCs of the HRP and DNR tends to add the minimum number of copies of a notification when the output queues of the target links are congested. The average notification delay in the DNR algorithm when the notification rate was 100K was 17.5\% less than the SNR algorithm. Similar improvements, when the rates were 80K, 60K messages per minute, were 17.7\%, and 21\%, respectively. On average, DNR algorithm stabilized the system 239 seconds before the SNR algorithm and generated only 0.3\% IMs more than the SNR algorithm and 58.6\% less than the BID-based routing algorithm. Analysis of the collected data indicates that the number of messages that had delivery delays less than 1 second in the DNR, SNR and BID-based routing algorithms are 53.8\%, 44.1\% and 31\%, respectively. We also conducted several experiments with HRP on each cluster and sending notifications with different rates. The performance difference between DNR and SNR decreased with an increase in the number of HRPs and decrease in their burst rates.  

\section{Conclusion}
This paper introduces an advertisement-based CPS system that offers inter-cluster dynamic routing of notifications. CPUG is used to define a structured cyclic overlay SCOT to generate subscription trees of the shortest lengths and eliminate cycles in notification routing. The main contribution of this paper is the IDR algorithm that uses a lightweight bit vector mechanism with cluster-based SCOT to provide dynamic routing of notifications between host clusters of interested subscribers. The algorithm does not require a global knowledge of an overlay topology and dynamic routing is done when congestion is detected in output queues of a broker. An implementation of the IDR and SNR algorithms in PADRES indicates that the IDR and SNR scale well with the number of publishers and subscribers.

\bibliographystyle{plain}
\bibliography{refs_p1}

\end{document}